\documentclass[a4paper,11pt]{article}
\pdfoutput=1 

\usepackage{jcappub} 
\usepackage{amsmath,graphicx,hyperref,bm,hyperref}
\usepackage{color}
\title{\boldmath The Third Order Scalar Induced Gravitational Waves}

\author[a,b,1]{Jing-Zhi Zhou,\note{Corresponding author.}}
\author[a,b]{Xukun Zhang,}
\author[a,b]{Qing-Hua Zhu,}
\author[a,b]{Zhe Chang}

\affiliation[a]{Institute of High Energy Physics, Chinese Academy of Sciences, Beijing 100049, China}
\affiliation[b]{University of Chinese Academy of Sciences, Beijing 100049, China}

\emailAdd{zhoujingzhi@ihep.ac.cn}
\emailAdd{zhangxukun@ihep.ac.cn}
\emailAdd{zhuqh@ihep.ac.cn}
\emailAdd{changz@ihep.ac.cn}

\begin{abstract}
	{Since the gravitational waves were detected by LIGO and Virgo, it has been promising that lots of information about the primordial Universe could be learned by further observations on stochastic gravitational waves background. The studies on gravitational waves induced by primordial curvature perturbations are of great interest.
	The aim of this paper is to investigate the third order induced gravitational waves. 
	Based on the theory of cosmological perturbations, the first order scalar induces the second order scalar, vector and tensor perturbations. At the next iteration, the first order scalar, the second order scalar, vector and tensor perturbations all induce the third order tensor perturbations. We present the two point function $\langle h^{\lambda,(3)}h^{\lambda',(3)} \rangle$ and corresponding energy density spectrum of the third order gravitational waves for a monochromatic primordial power spectrum.
	The shape of the energy density spectrum of the third order gravitational waves is different from that of the second order scalar induced gravitational waves. And it is found that the third order gravitational waves sourced by the second order scalar perturbations dominate the two point function $\langle h^{\lambda,(3)}h^{\lambda',(3)} \rangle$ and corresponding energy density spectrum of third order scalar induced gravitational waves.}
\end{abstract}

\begin{document}
\maketitle
\flushbottom

\section{Introduction}
Inflationary cosmology suggests that cosmological perturbations are originated from quantum fluctuations in causal contact at the early Universe. Since the gravitational waves were detected by LIGO and Virgo \cite{Abbott:2016blz}, it has been promising that the further observations on stochastic gravitational waves background could test inflationary models and might shed light on the quantum reality of the early Universe \cite{Sasaki:2018dmp,Arzoumanian:2015liz,Arzoumanian:2018saf,Arzoumanian:2020vkk}.

The cosmological perturbations are decomposed as scalar, vector and tensor perturbations based on helicity decomposition of metric perturbations. The cosmological perturbations generated at inflation epochs are known as primordial perturbations. So far, the measurements of cosmic microwave background  and large-scale structure indicate that the scalar parts dominate the primordial perturbations \cite{Akrami:2018odb,Aghanim:2018eyx}. Therefore, it is of great interest to study gravitational waves induced by the primordial scalar perturbations \cite{Ananda:2006af,Baumann:2007zm,Assadullahi:2009jc,Alabidi:2012ex,Guzzetti:2016mkm,Hwang:2017oxa,Caprini:2018mtu,Kohri:2018awv,Tomikawa:2019tvi,DeLuca:2019ufz,Inomata:2019yww,Yuan:2019fwv,Domenech:2019quo,Domenech:2020kqm,Pi:2020otn,Ali:2020sfw,Chang:2020iji,Chang:2020mky,Abe:2020sqb,Fumagalli:2020nvq,Fumagalli:2021cel,Espinosa:2018eve,Cai:2021yvq}, if there would be further directed observation on the gravitational wave background. Besides, the cosmic microwave background  and large-scale structure constrained the primordial spectrum on the pivot scale larger than $1$Mpc \cite{Akrami:2018odb}. It did not tell the story about perturbations that re-enter the horizon on small scales, which also contain lots of information about the early Universe. In this sense, it is necessary to apply to future space-based and ground-based detections of stochastic gravitational waves background on small scales \cite{amaroseoane2017laser,Luo:2015ght,Guo:2018npi,Arzoumanian:2020vkk}, and theoretically study cosmological perturbations from the early Universe, such as the induced gravitational waves.

The induced gravitational wave as a prediction of inflationary cosmology has been studied for many years \cite{Ananda:2006af,Baumann:2007zm,Matarrese:1993zf}.  
The energy density spectra of the induced gravitational waves in principle include the information about primordial black hole \cite{Saito:2008jc,Bugaev:2009kq,Saito:2009jt,Bugaev:2010bb,Byrnes:2018txb,Clesse:2018ogk,Cai:2019jah,Wang:2019kaf,Tada:2019amh,Yuan:2019udt,Carrilho:2019oqg,Lu:2019sti,Bartolo:2019zvb,Bhattacharya:2019bvk,Cai:2019bmk,Lin:2020goi,Ballesteros:2020qam,Inomata:2020lmk,Fumagalli:2020adf,Braglia:2020eai,Dalianis:2020cla,Yi:2020kmq,Green:2020jor,Ragavendra:2020sop,Bhaumik:2020dor,Kohri:2020qqd,Zhou:2020kkf,Domenech:2020ers,Papanikolaou:2020qtd,Inomata:2020xad,Ragavendra:2020vud,Yi:2020cut,Gao:2020tsa,Domenech:2020ssp,DeLuca:2020agl,Wang:2016ana,Li:2017ydy} and primordial non-Gaussianity \cite{Bartolo:2018qqn,Cai:2018dig,Unal:2018yaa,Ota:2020vfn,Yuan:2020iwf,Vagnozzi:2020gtf,Adshead:2021hnm}. Besides, recent studies on induced gravitational wave were also extended to gauge   issue \cite{Tomikawa:2019tvi,DeLuca:2019ufz,Inomata:2019yww,Yuan:2019fwv,Ali:2020sfw,Chang:2020iji,Chang:2020mky,Lu:2020diy}, epochs of the Universe \cite{Kohri:2018awv,Domenech:2020kqm,Domenech:2019quo} and modified gravity \cite{Ananda:2007xh}. 

As suggested in pioneers' study \cite{Yuan:2019udt}, the higher order induced gravitational waves have non-trivial results comparing with the second order induced gravitational waves, and play important role in constraints on primordial black hole. Because, the existence of primordial black holes generated at the early time may suggest a quiet large amplitude of primordial curvature perturbation at small scale. This makes the higher order perturbations considerable. In this paper, we also study induced gravitational waves to the third order in a radiation-dominated era. For scalar induced gravitational waves \cite{Baumann:2007zm,Yuan:2021qgz}, the source terms of the third order gravitational waves should contain not only the first order scalar perturbations, but also other three types of the second order perturbations, since the second order perturbations can also be induced by the first order scalar perturbations. It is found that the dominated parts of the power spectrum of the third order gravitational waves for a monochromatic primordial power spectrum come from the source term of the second order scalar perturbations. This is beyond previous study \cite{Yuan:2019udt,Yuan:2021qgz} that neglected all the second order perturbations.

This paper is organized as follows. In Sec.~\ref{sec:eh3}, we present the equation of motion of the third order induced gravitational waves. 
In Sec.~\ref{sec:eh30}, we solve the equation of motion of the third order induced gravitational waves. The explicit expressions of kernel functions are obtained. In Sec.~\ref{sec:Ph}, the energy density soectrum of the third order induced gravitational waves is studied. Finally, the conclusions and discussions are summarized in Sec.~\ref{sec:CaD}.

\section{Equations of motion of third order induced gravitational waves}\label{sec:eh3}
In flat Friedmann-Robertson-Walker (FRW) spacetime, the background metric is given by
\begin{equation}
	g_{\mu \nu}^{(0)} \mathrm{d} x^{\mu} \mathrm{d} x^{\nu}=a^{2}(\eta)\left(-\mathrm{d} \eta^{2}+\delta_{i j} \mathrm{d} x^{i} \mathrm{d} x^{j}\right) \ ,
\end{equation}
where $\eta$ is the conformal time. The third order metric perturbations in Newtonian gauge take the form of
\begin{equation}
	\begin{aligned}
		\mathrm{d} s^{2}&=-a^{2}\Bigg[\left(1+2 \phi^{(1)}+ \phi^{(2)}\right) \mathrm{d} \eta^{2}+ V_i^{(2)} \mathrm{d} \eta \mathrm{d} x^{i} \\
		&+\left(\left(1-2 \psi^{(1)}- \psi^{(2)}\right) \delta_{i j}+\frac{1}{2} h_{i j}^{(2)}+\frac{1}{6} h_{i j}^{(3)}\right)\mathrm{d} x^{i} \mathrm{d} x^{j}\Bigg] \ ,
	\end{aligned}
\end{equation}
where $\phi^{(n)}$ and $\psi^{(n)}$$(n=1,2)$ are the $n$-order scalar perturbations,  $V^{(2)}_i$ is the second order vector perturbation, and $h_{i j}^{(n)}$ $(n=2,3)$ are the $n$-order tensor perturbations. 
The Einstein equation $G_{\mu\nu}=\kappa T_{\mu\nu}$ of the third order perturbations are evaluated by using \texttt{xPand} package \cite{Pitrou:2013hga}.
The space-space part of the third order perturbed equation is presented as follows,
\begin{equation}\label{eq:eq}
	\begin{aligned}
		&\frac{1}{12}h^{(3)''}_{ij}+\frac{1}{6}h^{(3)'}_{ij}\mathcal{H}-\frac{1}{6}h^{(3)}_{ij}\mathcal{H}^2-\frac{1}{3}h^{(3)}_{ij}\mathcal{H}^{'}-\frac{1}{18}\kappa a^2h^{(3)}_{ij}\rho^{(0)}-\frac{1}{12}\Delta h^{(3)}_{ij}-\frac{2}{3}\kappa a^2 V_j^{(2)}u^{(1)}_i\rho^{(0)}\\
		&-\frac{2}{3}\kappa a^2 V_i^{(2)}u^{(1)}_j\rho^{(0)}-\frac{2}{3}\kappa a^2u^{(1)}_iu^{(2)}_j\rho^{(0)}-\frac{2}{3}\kappa a^2u^{(1)}_ju^{(2)}_i\rho^{(0)}-\frac{1}{6}\kappa a^2h^{(2)}_{ij}\rho^{(1)}-\frac{4}{3}\kappa a^2u^{(1)}_iu^{(1)}_j\rho^{(1)}\\
		&-h^{(2)'}_{ij}\mathcal{H}\phi^{(1)}+h^{(2)}_{ij}\mathcal{H}^2\phi^{(1)}+2h^{(2)}_{ij}\mathcal{H}^{'}\phi^{(1)}+\frac{16}{3}\kappa a^2u^{(1)}_iu^{(1)}_j\rho^{(0)}\phi^{(1)}+4h^{(2)}_{ij}\mathcal{H}\phi^{(1)'}+\frac{3}{2}h^{(2)}_{ij}\phi^{(1)''}\\
		&-\frac{1}{2}h^{(2)}_{ij}\Delta \phi^{(1)}+\frac{1}{2}h^{b(2)}_{i}\partial_b\partial_j\phi^{(1)}+\frac{1}{2}h^{b(2)}_{j}\partial_b\partial_i\phi^{(1)}-\partial_bh^{(2)}_{ij}\partial^b\phi^{(1)}+\mathcal{H}\phi^{(1)}\partial_iV^{(2)}_j+\frac{1}{2}\phi^{(1)'}\partial_iV^{(2)}_j\\
		&+\frac{1}{2}\partial^b\phi^{(1)}\partial_ih^{(2)}_{jb}-\frac{1}{2}V^{(2)'}_j\partial_i\phi^{(1)}-V^{(2)}_j\mathcal{H}\partial_i\phi^{(1)}-\frac{1}{2}V^{(2)}_j\partial_i\phi^{(1)'}+\mathcal{H}\phi^{(1)}\partial_jV^{(2)}_i+\frac{1}{2}\phi^{(1)'}\partial_jV^{(2)}_i\\
		&+\frac{1}{2}\phi^{(1)}\partial_jV^{(2)}_i+\frac{1}{2}\partial^b\phi^{(1)}\partial_jh^{(2)}_{ib}-\frac{1}{2}V^{(2)'}_i\partial_j\phi^{(1)}-V^{(2)}_i\mathcal{H}\partial_j\phi^{(1)}-\frac{1}{2}V^{(2)}_i\partial_j\phi^{(1)'}+8\phi^{(1)}\partial_i\phi^{(1)}\partial_j\phi^{(1)}\\
		&+\partial_i\psi^{(2)}\partial_j\phi^{(1)}+\partial_i\phi^{(1)}\partial_j\psi^{(2)}+\phi^{(2)}\partial_i\partial_j\phi^{(1)}+\phi^{(1)}\partial_i\partial_j\phi^{(2)}+\psi^{(2)}\partial_i\partial_j\phi^{(1)}+\phi^{(1)}\partial_i\partial_j\psi^{(2)} \\
		&-\frac{1}{2}\phi^{(1)}\Delta h^{(2)}_{ij}-\frac{1}{2}h^{(2)''}_{ij}\phi^{(1)}+\frac{1}{2}\phi^{(1)}\partial_iV^{(2)'}_j=0
		\ ,
	\end{aligned}
\end{equation}
where $u^{(1)}_i$ and $u^{(2)}_i$ are first and second order transverse part of three dimensional velocity in  energy-momentum tensor. We have set $\psi^{(1)}=\phi^{(1)}$, according to the equation of motion of first order scalar perturbations. In order to obtain equations of the induced gravitational waves, we express $\rho^{(0)}$, $\rho^{(1)}$, $u^{(1)}$, $u^{(2)}$, and $\mathcal{H}^{'}$ in Eq.~(\ref{eq:eq}) in terms of $\mathcal{H}$, the first order perturbations, and the second order perturbations. Namely, we substitute Eqs.~(\ref{eq:e1})--(\ref{eq:e4}) into Eq.~(\ref{eq:eq}),
\begin{equation}\label{eq:e1}
	\begin{aligned}
		\mathcal{H}^{'}=-\mathcal{H}^2 \ \ , \ \ \rho^{(0)}=\frac{3\mathcal{H}^{2}}{\kappa a^2} \ ,
	\end{aligned}
\end{equation}
\begin{equation}\label{eq:e2}
	\begin{aligned}
		\rho^{(1)}=\frac{-6\mathcal{H}\left(\mathcal{H}\phi^{(1)}+\psi^{(1)'}\right)+2\Delta \psi^{(1)}}{\kappa a^2} \ \ , \ \ u^{(1)}_i=-\frac{\left(\mathcal{H}\partial_i\phi^{(1)}+\partial_i\psi^{(1)'}\right)}{2\mathcal{H}^2} \ ,
	\end{aligned}
\end{equation}
\begin{equation}\label{eq:e3}
	\begin{aligned}
		\rho^{(2)}&=\frac{1}{4\kappa a^2\mathcal{H}}\Bigg[ 2\mathcal{H}\Bigg( 12\mathcal{H}^3\left( 4(\phi^{(1)})^2-\phi^{(2)}\right)-12\mathcal{H}^2\psi^{(2)'}-8\partial_b\phi^{(1)'} \partial^b\phi^{(1)}\\
		&+\mathcal{H}\left(12(\phi^{(1)'})^2+32\phi^{(1)}\Delta\phi^{(1)}+4\Delta\psi^{(2)}+8\partial_b\phi^{(1)}\partial^b\phi^{(1)}\right)   \Bigg)-8\partial_b\phi^{(1)'}\partial^b\phi^{(1)'}\Bigg] \ ,
	\end{aligned}
\end{equation}
\begin{equation}\label{eq:e4}
	\begin{aligned}
		u^{(2)}_i&=-V^{(2)}_i+\frac{1}{32\mathcal{H}^4}\Bigg[\frac{64}{3}\Delta\phi^{(1)}\partial_i\phi^{(1)'}-\frac{64}{3}\mathcal{H}\left(-\Delta \phi^{(1)}\partial_i\phi^{(1)}+3\phi^{(1)'}\partial_i\phi^{(1)'} \right)\\ 
		&+3\mathcal{H}^2\left(\frac{4}{3}\Delta V_i^{(2)}-32\phi^{(1)'}\partial_i\phi^{(1)}-\frac{160}{3} \phi^{(1)}\partial_i\phi^{(1)'}-\frac{16}{3}\partial_i\psi^{(2)'}\right)\\
		&+12\mathcal{H}^3\left(-\frac{8}{3}\phi^{(1)}\partial_i\phi^{(1)}-\frac{4}{3}\partial_i\phi^{(2)}\right)\Bigg] \ .
	\end{aligned}
\end{equation}
Therefore, the equation of motion of the third order induced gravitational waves is rewritten as follows,
\begin{equation}\label{eq:h3}
	h_{i j}^{(3)''}(\eta,\mathbf{x})+2 \mathcal{H}  h_{i j}^{(3)'}(\eta,\mathbf{x})-\Delta h_{i j}^{(3)}(\eta,\mathbf{x})=-12 \Lambda_{i j}^{l m} \mathcal{S}^{(3)}_{l m}(\eta,\mathbf{x}) \ .
\end{equation}
The expression of transverse and traceless operator is
\begin{equation}
	\begin{aligned}
		\Lambda_{i j}^{l m}&=\mathcal{T}_{i}^{l} \mathcal{T}_{j}^{m}-\frac{1}{2} \mathcal{T}_{i j} \mathcal{T}^{l m}~,
	\end{aligned}
\end{equation}
where the expression of traceless operator is $\mathcal{T}_{i}^{l}=\delta_{i}^{l}-\partial^{l} \Delta^{-1} \partial_{i}$. The explicit details of the decomposed operators are shown in Appendix~\ref{sec:A}.
For illustration, the source term $S_{lm}^{(3)}(\eta,\mathbf{x})$ can be divided into four parts,
\begin{equation}
	S_{lm}^{(3)}(\eta,\mathbf{x})=S_{lm,1}^{(3)}(\eta,\mathbf{x})+S_{lm,2}^{(3)}(\eta,\mathbf{x})+S_{lm,3}^{(3)}(\eta,\mathbf{x})+S_{lm,4}^{(3)}(\eta,\mathbf{x}) \ ,
\end{equation}
where $S_{lm,1}^{(3)}(\eta,\mathbf{x})$ is composed of the first order scalar perturbation $\phi^{(1)}$, 
\begin{equation}
	\begin{aligned}
		S_{lm,1}^{(3)}(\eta,\mathbf{x})&=12\phi^{(1)}\partial_l\phi^{(1)}\partial_m\phi^{(1)}-\frac{4}{\mathcal{H}}\phi^{(1)'}\partial_l\phi^{(1)}\partial_m\phi^{(1)}+\frac{2}{3\mathcal{H}^2}\Delta\phi^{(1)}\partial_l\phi^{(1)}\partial_m\phi^{(1)}\\
		&+\frac{2}{3\mathcal{H}^4}\Delta\phi^{(1)}\partial_l\phi^{(1)'}\partial_m\phi^{(1)'}-\frac{3}{\mathcal{H}^2}\phi^{(1)'}\partial_l\phi^{(1)'}\partial_m\phi^{(1)}-\frac{3}{\mathcal{H}^2}\phi^{(1)'}\partial_m\phi^{(1)'}\partial_l\phi^{(1)}\\
		&+\frac{2}{3\mathcal{H}^3}\Delta\phi^{(1)}\partial_l\phi^{(1)'}\partial_m\phi^{(1)}+\frac{2}{3\mathcal{H}^3}\Delta\phi^{(1)}\partial_m\phi^{(1)'}\partial_l\phi^{(1)}\\
		&-\frac{2}{\mathcal{H}^3}\phi^{(1)'}\partial_l\phi^{(1)'}\partial_m\phi^{(1)'}-\frac{4}{\mathcal{H}^2}\phi^{(1)}\partial_l\phi^{(1)'}\partial_m\phi^{(1)'} \ .
	\end{aligned}\label{Z5}
\end{equation}
The source term $S_{lm,2}^{(3)}(\eta,\mathbf{x})$ is composed of the first order scalar perturbation $\phi^{(1)}$ and the second order tensor perturbation $h_{lm}^{(2)}$. 
The expression of $S_{lm,2}^{(3)}(\eta,\mathbf{x})$ is shown to be
\begin{equation}
	\begin{aligned}
		S_{lm,2}^{(3)}(\eta,\mathbf{x})&=-\frac{1}{2}\phi^{(1)}\left( h_{lm}^{(2)''}+2 \mathcal{H}  h_{lm}^{(2)'}-\Delta h_{lm}^{(2)}\right)-\phi^{(1)}\Delta h_{lm}^{(2)}-\phi^{(1)'}\mathcal{H}h_{lm}^{(2)}-\frac{1}{3}\Delta \phi^{(1)}h_{lm}^{(2)}\\
		&-\partial^b \phi^{(1)}\partial_b h_{lm}^{(2)} \ .
	\end{aligned} 
\end{equation}
The source term $S_{lm,3}^{(3)}(\eta,\mathbf{x})$ is composed of the first order scalar perturbation $\phi^{(1)}$ and the second order vector perturbation $V_l^{(2)}$, namely, 
\begin{equation}
	\begin{aligned}
		S_{lm,3}^{(3)}(\eta,\mathbf{x})&=\phi^{(1)}\partial_l\left(V_m^{(2)'}+2 \mathcal{H}V_m^{(2)} \right)+\phi^{(1)}\partial_m\left(V_l^{(2)'}+2 \mathcal{H}V_l^{(2)} \right)+\phi^{(1)'}\left(\partial_lV_m^{(2)}+\partial_mV_l^{(2)}\right)\\
		&-\frac{\phi^{(1)}}{8\mathcal{H}}\left(\partial_m\Delta V_l^{(2)}+\partial_l\Delta V_m^{(2)}\right)-\frac{\phi^{(1)'}}{8\mathcal{H}^2}\left(\partial_m\Delta V_l^{(2)}+\partial_l\Delta V_m^{(2)}\right) \ .
	\end{aligned}
\end{equation}
And the source term $S_{lm,4}^{(3)}(\eta,\mathbf{x})$ is composed of the first order scalar perturbation $\phi^{(1)}$ and the second order scalar perturbations $\phi^{(2)}$ and $\psi^{(2)}$, 
\begin{equation}
	\begin{aligned}
		S_{lm,4}^{(3)}(\eta,\mathbf{x})&=\frac{1}{\mathcal{H}}\left(\phi^{(1)}\partial_l\partial_m\psi^{(2)'}\right)+\frac{1}{\mathcal{H}}\left(\phi^{(1)'}\partial_l\partial_m\phi^{(2)}\right)+\frac{1}{\mathcal{H}^2}\left(\phi^{(1)'}\partial_l\partial_m\psi^{(2)'}\right)\\
		&+3\left(\phi^{(1)}\partial_l\partial_m\phi^{(2)}\right) \ .
	\end{aligned}\label{Z8}
\end{equation}

In this paper, the second order perturbations $h_{lm}^{(2)}$,  $V_l^{(2)}$ and $\phi^{(2)}$ and $\psi^{(2)}$ are induced by the first order scalar perturbations $\psi^{(1)}(=\phi^{(1)})$. The equations of motion of the second order perturbations are shown in Appendix~\ref{sec:Lower}. In this case, 
the third gravitational waves are all attributed to the first order primordial curvature perturbations.
In the following, we will solve the motion of equation for the third order gravitational waves $h^{(3)}_{i j}$ based on the expression of Eq.~(\ref{eq:h3}).

\section{Kernel function of third order induced gravitational waves}\label{sec:eh30}
In order to solve the equations of motion of third order induced gravitational waves, we rewrite Eq.~(\ref{eq:h3})  in momentum space as
\begin{equation}\label{eq:h3m}
	\begin{aligned}
		h^{\lambda,(3)''}(\eta,\mathbf{k})+2 \mathcal{H}h^{\lambda,(3)'}(\eta,\mathbf{k})+k^{2} h^{\lambda,(3)}(\eta,\mathbf{k}) = \sum^{4}_{i=1} 12\mathcal{S}_i^{\lambda,(3)}(\eta,\mathbf{k}) ~,
	\end{aligned}
\end{equation}
where $h^{\lambda,(3)}(\eta,\mathbf{k})=\varepsilon^{\lambda, ij}(\mathbf{k})h_{ij}^{(3)}(\eta,\mathbf{k})$ and  $S^{\lambda,(3)}(\eta,\mathbf{k})=-\varepsilon^{\lambda, lm}(\mathbf{k})S_{lm}^{(3)}(\eta,\mathbf{k})$. The $\varepsilon_{i j}^{\lambda}(\mathbf{k})$ is polarization tensor, which satisfies $\varepsilon_{i j}^{\lambda}(\mathbf{k}) \varepsilon^{\bar{\lambda}, ij}(\mathbf{k})=\delta^{\lambda \bar{\lambda}}$ and $\delta_{\lambda \bar{\lambda}} \varepsilon_{i j}^{\lambda}(\mathbf{k}) \varepsilon^{\bar{\lambda}, l m}(\mathbf{k})=\Lambda_{i j}^{l m}(\mathbf{k})$.
Unlike the second order tensor perturbation, the third order induced gravitational waves have four types of source terms $S_i^{\lambda,(3)}(\eta,\mathbf{k}),(i=1,2,3,4)$ (Eqs.~(\ref{Z5})--(\ref{Z8})).
As mentioned, we consider that the second order perturbations $h_{ij}^{(2)}$, $V_l^{(2)}$, $\phi^{(2)}$, and $\psi^{(2)}$ are all induced by the first order scalar perturbations $\phi^{(1)}$ and $\psi^{(1)}$. It is convenient to rewrite the first order scalar perturbations in the form of
\begin{equation}
	\psi(\eta,\mathbf{p}) = \phi(\eta,\mathbf{p}) = \Phi_{\mathbf{p}} T_\phi(p \eta)~,
\end{equation}
where $\Phi_{\mathbf{p}}$ is initial value originated from primordial curvature perturbation, and transfer function $T_{\phi}(y)=\frac{9}{y^{2}}\left(\frac{\sqrt{3}}{y} \sin \left(\frac{y}{\sqrt{3}}\right)-\cos \left(\frac{y}{\sqrt{3}}\right)\right) $.
Therefore, we can rewrite the source terms $S_i^{\lambda,(3)}(\eta,\mathbf{k})$ in terms of initial value of the first order scalar perturbations $\Phi_{\mathbf{p}}$ in momentum space, i.e.,
\begin{eqnarray}
	S_1^{\lambda,(3)}(\eta,\mathbf{k})&=&\int\frac{d^3p}{(2\pi)^{3/2}}\int\frac{d^3q}{(2\pi)^{3/2}}\varepsilon^{\lambda,lm}(\mathbf{k})(p_l-q_l)q_m\Phi_{\mathbf{k}-\mathbf{p}} \Phi_{\mathbf{p}-\mathbf{q}} \Phi_{\mathbf{q}}\nonumber\\
	& &\times
	f_1^{(3)}(|\mathbf{k}-\mathbf{p}|,|\mathbf{p}-\mathbf{q}|,\mathbf{q},\eta) \ , \label{eq:S31}\\
	S_2^{\lambda,(3)}(\eta,\mathbf{k})&=&\int\frac{d^3p}{(2\pi)^{3/2}}\int\frac{d^3q}{(2\pi)^{3/2}}\varepsilon^{\lambda, lm}(\mathbf{k})\Lambda^{r s}_{l m}(\textbf{p})q_rq_s\Phi_{\mathbf{k}-\mathbf{p}} \Phi_{\mathbf{p}-\mathbf{q}} \Phi_{\mathbf{q}}\nonumber\\
	& &\times
	f^{(3)}_2(|\mathbf{k}-\mathbf{p}|,|\mathbf{p}-\mathbf{q}|,\mathbf{q},\eta) \ ,  \label{eq:S32} \\
	S_3^{\lambda,(3)}(\eta,\mathbf{k})&=&\int\frac{d^3p}{(2\pi)^{3/2}}\int\frac{d^3q}{(2\pi)^{3/2}}\varepsilon^{\lambda,lm}(\mathbf{k})\left(\mathcal{T}^s_m(\textbf{p}) p_l+\mathcal{T}^s_l(\textbf{p})p_m\right)\frac{p^s}{p^2}q_rq_s\Phi_{\mathbf{k}-\mathbf{p}} \Phi_{\mathbf{p}-\mathbf{q}} \Phi_{\mathbf{q}} \nonumber\\
	& &\times f^{(3)}_3(|\mathbf{k}-\mathbf{p}|,|\mathbf{p}-\mathbf{q}|,\mathbf{q},\eta) \ ,\label{eq:S33} \\
	S_4^{\lambda,(3)}(\eta,\mathbf{k})&=&\int\frac{d^3p}{(2\pi)^{3/2}}\int\frac{d^3q}{(2\pi)^{3/2}}\varepsilon^{\lambda,lm}(\mathbf{k})p_lp_m\Phi_{\mathbf{k}-\mathbf{p}} \Phi_{\mathbf{p}-\mathbf{q}} \Phi_{\mathbf{q}}\nonumber\\
	& &\times
	f^{(3)}_4(|\mathbf{k}-\mathbf{p}|,|\mathbf{p}-\mathbf{q}|,\mathbf{q},\eta) \ , \label{eq:S34}
\end{eqnarray}
and the expression of the transfer functions $f_i^{(3)}(|\mathbf{k}-\mathbf{p}|,|\mathbf{p}-\mathbf{q}|,\mathbf{q},\eta)$ 
are shown as follows,
\begin{equation}
	\begin{aligned}
		f_1^{(3)}(u,\bar{u},\bar{v},x,y)&=12T_{\phi}(ux) T_{\phi}(\bar{u}y) T_{\phi}(\bar{v}y)-4ux\frac{d}{d(ux)}T_{\phi}(ux) T_{\phi}(\bar{u}y) T_{\phi}(\bar{v}y)\\
		&-\frac{2u^2x^2}{3}T_{\phi}(ux) T_{\phi}(\bar{u}y) T_{\phi}(\bar{v}y)-6\bar{u}uxy\frac{d}{d(ux)}T_{\phi}(ux) \frac{d}{d(\bar{u}y)}T_{\phi}(\bar{u}y) T_{\phi}(\bar{v}y)\\
		&-\frac{4u^2\bar{u}x^2y}{3}T_{\phi}(ux)\frac{d}{d(\bar{u}y)} T_{\phi}(\bar{u}y) T_{\phi}(\bar{v}y)-4\bar{u}\bar{v}y^2T_{\phi}(ux) \frac{d}{d(\bar{u}y)}T_{\phi}(\bar{u}y) \frac{d}{d(\bar{v}y)}T_{\phi}(\bar{v}y)\\
		&-2\bar{u}\bar{v}uy^2x\frac{d}{d(ux)}T_{\phi}(ux) \frac{d}{d(\bar{u}y)}T_{\phi}(\bar{u}y) \frac{d}{d(\bar{v}y)}T_{\phi}(\bar{v}y)\\
		&-\frac{2u^2\bar{u}\bar{v}x^2y^2}{3}T_{\phi}(ux) \frac{d}{d(\bar{u}y)}T_{\phi}(\bar{u}y) \frac{d}{d(\bar{v}y)}T_{\phi}(\bar{v}y)\ ,
	\end{aligned}
\end{equation}

\begin{equation}
	\begin{aligned}
		f_2^{(3)}(u,\bar{u},\bar{v},x,y)=&-6T_{\phi}(ux) T_{\phi}(\bar{u}y) T_{\phi}(\bar{v}y)-4\bar{u}yT_{\phi}(ux) \frac{d}{d(\bar{u}y)}T_{\phi}(\bar{u}y) T_{\phi}(\bar{v}y)\\
		&-T_{\phi}(ux)p^2I_{h}^{(2)}(\bar{u},\bar{v},y)-2\bar{u}\bar{v}y^2T_{\phi}(ux) \frac{d}{d(\bar{u}y)}T_{\phi}(\bar{u}y) \frac{d}{d(\bar{v}y)}T_{\phi}(\bar{v}y)\\
		&+\frac{u}{v^2x}\frac{d}{d(ux)}T_{\phi}(ux)p^2I_{h}^{(2)}(\bar{u},\bar{v},y)-\frac{u^2}{3v^2} T_{\phi}(ux)p^2I_{h}^{(2)}(\bar{u},\bar{v},y)\\
		&-\frac{1-u^2-v^2}{2v^2} T_{\phi}(ux)p^2 I_{h}^{(2)}(\bar{u},\bar{v},y) \ ,
	\end{aligned}\label{Z15}
\end{equation}

\begin{equation}
	\begin{aligned}
		f_3^{(3)}(u,\bar{u},\bar{v},x,y)&=\frac{u}{v}\frac{d}{d(ux)}T_{\phi}(ux)pI_V^{(2)}(\bar{u},\bar{v},y)+\frac{y}{8}T_{\phi}(ux)pI_V^{(2)}(\bar{u},\bar{v},y)\\
		&+\frac{xuy}{8}\frac{d}{d(ux)}T_{\phi}(ux)pI_V^{(2)}(\bar{u},\bar{v},y)-4T_{\phi}(ux) T_{\phi}(\bar{u}y) T_{\phi}(\bar{v}y)\\
		&+8\bar{u}y T_{\phi}(ux) \frac{d}{d(\bar{u}y)}T_{\phi}(\bar{u}y) T_{\phi}(\bar{v}y)+16T_{\phi}(ux) T_{\phi}(\bar{u}y) T_{\phi}(\bar{v}y)\\
		&+4\bar{u}\bar{v}y^2T_{\phi}(ux) \frac{d}{d(\bar{u}y)}T_{\phi}(\bar{u}y) \frac{d}{d(\bar{v}y)}T_{\phi}(\bar{v}y) \ ,
	\end{aligned}
\end{equation}

\begin{equation}
	\begin{aligned}
		f_4^{(3)}(u,\bar{u},\bar{v},\eta)=&y\left(T_{\phi}(ux)\frac{\partial}{\partial y}I^{(2)}_{\psi}(\bar{u},\bar{v},y)\right)+uxy\left(\frac{d}{d(ux)}T_{\phi}(ux)\frac{\partial}{\partial y}I^{(2)}_{\psi}(\bar{u},\bar{v},y)\right)\\
		&+ux\left(\frac{d}{d(ux)}T_{\phi}(ux)(I^{(2)}_{\psi}(\bar{u},\bar{v},y)+f^{(2)}_{\phi}(\bar{u},\bar{v},y))\right)\\
		&+3\left(T_{\phi}(ux)(I^{(2)}_{\psi}(\bar{u},\bar{v},y)+f^{(2)}_{\phi}(\bar{u},\bar{v},y))\right) \ ,
	\end{aligned} \label{Z18}
\end{equation}
where we have set $|k-p|=uk$, $p=vk$, $|p-q|=\bar{u}p$, $q=\bar{v}p$, $x=k\eta$, and $y=p\eta$ for illustration. Indicated by Eqs.~(\ref{Z15})--(\ref{Z18}), we have to calculate the kernel function of the second order scalar, vector, tensor perturbations induced by the first order scalar perturbations $\big(I^{(2)}_{\psi}(\bar{u},\bar{v},y)$, $I^{(2)}_{\phi}(\bar{u},\bar{v},y)$, $I^{(2)}_{V}(\bar{u},\bar{v},y)$, and $I^{(2)}_{h}(\bar{u},\bar{v},y)\big)$. The kernel functions of the second order perturbations are shown in Appendix~\ref{sec:Lower}.

\begin{figure}
	\includegraphics[scale=0.8]{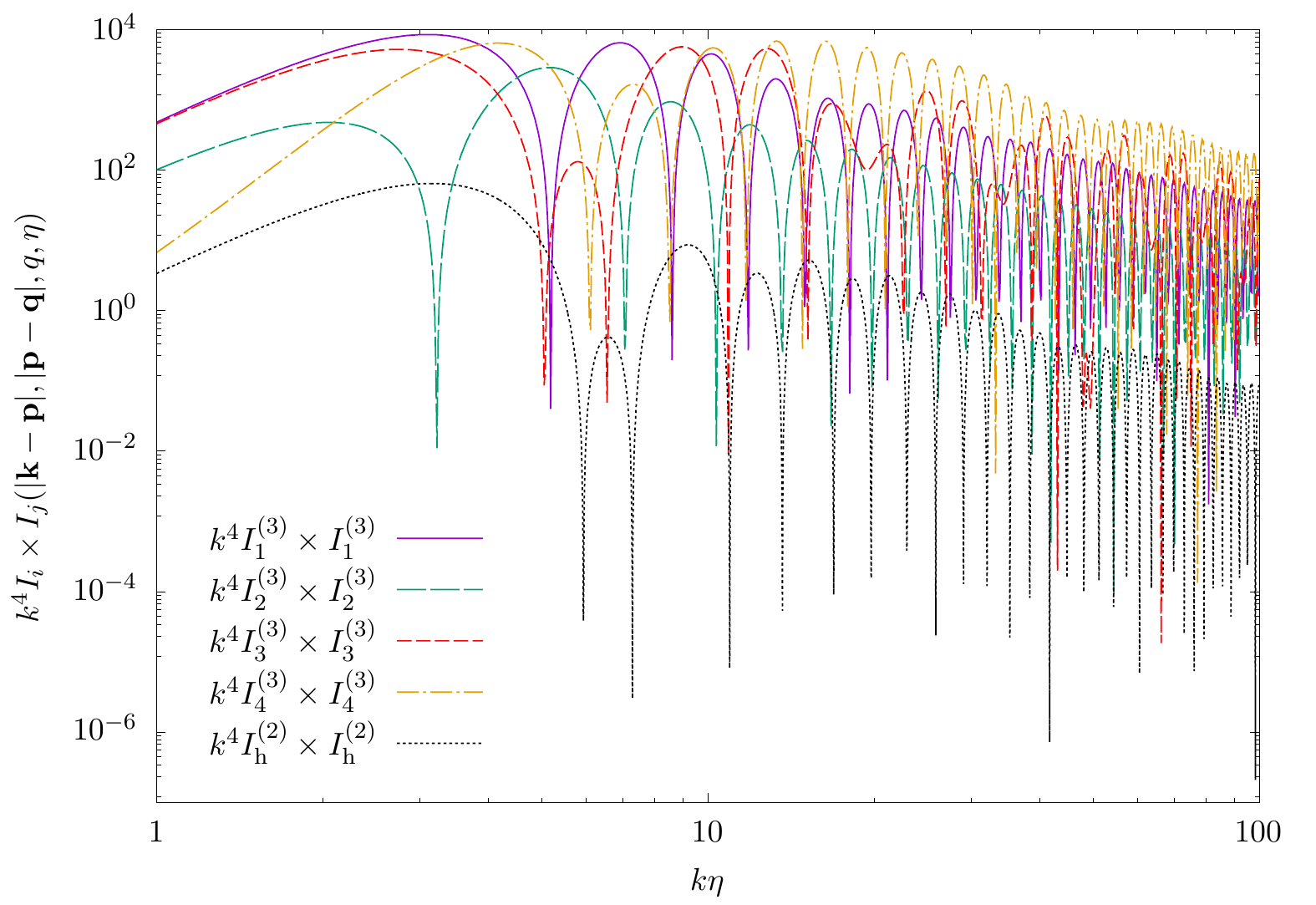}
	\caption{The squares of four types of the kernel functions of the third order tensor perturbations and the second order tensor perturbation. Here we have set $|{\bf k}-{\bf p}|=p=k$ and $|{\bf p}-{\bf q}|=q=p$.}\label{fig:3O_T_Ii_Ij1}
\end{figure}

\begin{figure}
	\includegraphics[scale=0.8]{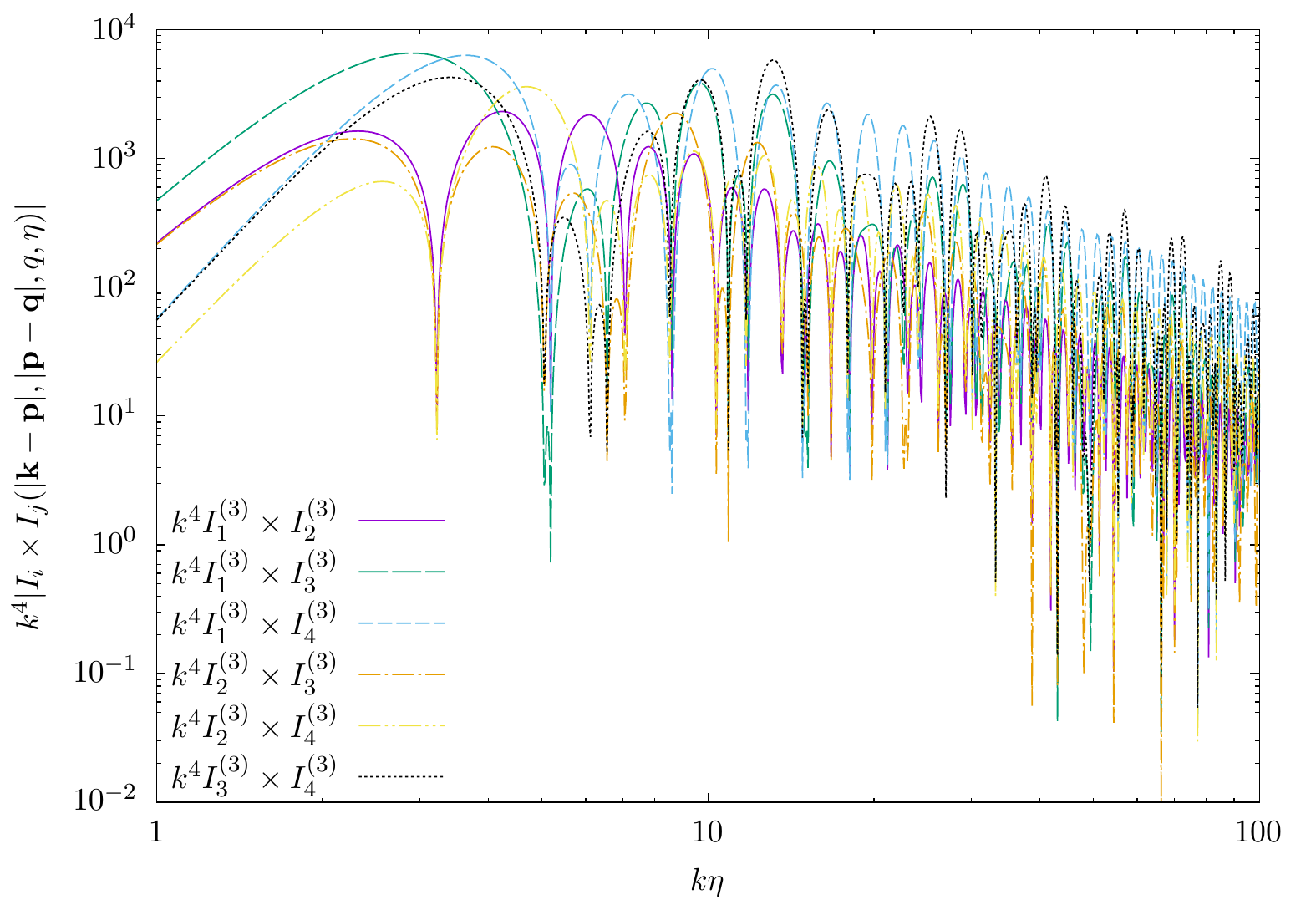}
	\caption{The products $k^4 |I_i\times I_j|\ (i\ne j)$ of the kernel functions of the third order tensor perturbations. Here we have set $|{\bf k}-{\bf p}|=p=k$ and $|{\bf p}-{\bf q}|=q=p$.}\label{fig:3O_T_Ii_Ij2}
\end{figure}

Corresponding to different types of the source terms in Eqs.~(\ref{eq:S31})--(\ref{eq:S34}), the third order gravitational wave can be divided into four types of parts,
\begin{equation}\label{eq:h30}
	\begin{aligned}
		h^{\lambda,(3)}(\eta,\mathbf{k})&=&h_1^{\lambda,(3)}(\eta,\mathbf{k})+h_2^{\lambda,(3)}(\eta,\mathbf{k})+h_3^{\lambda,(3)}(\eta,\mathbf{k})+h_4^{\lambda,(3)}(\eta,\mathbf{k}) \ ,
	\end{aligned}
\end{equation}
where $h_i^{\lambda,(3)},(i=1,2,3,4)$ can be defined with kernel functions $I^{(3)}_i(|\mathbf{k}-\mathbf{p}|,|\mathbf{p}-\mathbf{q}|,\mathbf{q},\eta)$, namely,
\begin{eqnarray}
	h_1^{\lambda,(3)}(\eta,\mathbf{k})&=&\int\frac{d^3p}{(2\pi)^{3/2}}\int\frac{d^3q}{(2\pi)^{3/2}}\varepsilon^{\lambda,lm}(\mathbf{k})(p_l-q_l)q_m\Phi_{\mathbf{k}-\mathbf{p}} \Phi_{\mathbf{p}-\mathbf{q}} \Phi_{\mathbf{q}} \nonumber\\
	& &\times
	I_1^{(3)}(|\mathbf{k}-\mathbf{p}|,|\mathbf{p}-\mathbf{q}|,\mathbf{q},\eta) \ ,\label{eq:h31}\\
	h_2^{\lambda,(3)}(\eta,\mathbf{k})&=&\int\frac{d^3p}{(2\pi)^{3/2}}\int\frac{d^3q}{(2\pi)^{3/2}}\varepsilon^{\lambda, lm}(\mathbf{k})\Lambda_{lm}^{ rs}(\mathbf{p})q_rq_s\Phi_{\mathbf{k}-\mathbf{p}} \Phi_{\mathbf{p}-\mathbf{q}} \Phi_{\mathbf{q}} \nonumber\\
	& &\times
	I^{(3)}_2(|\mathbf{k}-\mathbf{p}|,|\mathbf{p}-\mathbf{q}|,\mathbf{q},\eta) \ ,\label{eq:h32}\\
	h_3^{\lambda,(3)}(\eta,\mathbf{k})&=&\int\frac{d^3p}{(2\pi)^{3/2}}\int\frac{d^3q}{(2\pi)^{3/2}}\varepsilon^{\lambda,lm}(\mathbf{k})\left(\mathcal{T}^s_m(\textbf{p}) p_l+\mathcal{T}^s_l(\textbf{p})p_m\right)\frac{p^s}{p^2}q_rq_s\Phi_{\mathbf{k}-\mathbf{p}} \Phi_{\mathbf{p}-\mathbf{q}} \Phi_{\mathbf{q}} \nonumber\\
	& &\times I^{(3)}_3(|\mathbf{k}-\mathbf{p}|,|\mathbf{p}-\mathbf{q}|,\mathbf{q},\eta) \ ,\label{eq:h33}\\
	h_4^{\lambda,(3)}(\eta,\mathbf{k})&=&\int\frac{d^3p}{(2\pi)^{3/2}}\int\frac{d^3q}{(2\pi)^{3/2}}\varepsilon^{\lambda,lm}(\mathbf{k})p_lp_m\Phi_{\mathbf{k}-\mathbf{p}} \Phi_{\mathbf{p}-\mathbf{q}} \Phi_{\mathbf{q}} \nonumber\\
	& &\times
	I^{(3)}_4(|\mathbf{k}-\mathbf{p}|,|\mathbf{p}-\mathbf{q}|,\mathbf{q},\eta) \ .\label{eq:h34}
\end{eqnarray}
 Substituting Eqs.~(\ref{eq:S31})--(\ref{eq:S34}) and (\ref{eq:h31})--(\ref{eq:h34}) into Eq.~(\ref{eq:h3m}), we obtain the equations of motion of kernel functions $I_i^{(3)}(u,\bar{u},\bar{v},x)$ in the form of 
\begin{equation}
	\begin{aligned}
		I_i^{(3)''}(u,\bar{u},\bar{v},x)+2 \mathcal{H}I_i^{(3)'}(u,\bar{u},\bar{v},x)+k^{2} I_i^{(3)}(u,\bar{u},\bar{v},x)=12f_i^{(3)}(u,\bar{u},\bar{v},x) \ , \ (i=1,2,3,4) \ .
	\end{aligned}
\end{equation}
The solutions of kernel functions can be expressed as
\begin{equation}\label{eq:kerneli}
	\begin{aligned}
		I_{i}^{(3)}(u,\bar{u},\bar{v},x)=\frac{12}{k^{2}} \int_{0}^{x} \mathrm{d}\bar{x} \left(\frac{\bar{x}}{x} \sin(x-\bar{x}) f_{i}^{(3)}(u,\bar{u},\bar{v},\bar{x})\right) \ , \ (i=1,2,3,4) \ .
	\end{aligned}
\end{equation}
In this sense, the kernel functions $I^{(3)}_i(|\mathbf{k}-\mathbf{p}|,|\mathbf{p}-\mathbf{q}|,\mathbf{q},\eta)$ include the information of the corresponding source terms.
Since the expression of power spectrum is composed of product of kernel functions $I_i^{(3)}$, we here present the  $I_i(|\mathbf{k}-\mathbf{p}|,|\mathbf{p}-\mathbf{q}|,\mathbf{q},\eta)I_j(|\mathbf{k}-\mathbf{p}|,|\mathbf{p}-\mathbf{q}|,\mathbf{q},\eta), (i,j=1,2,3,4)$ as function of $k \eta$ in Fig.~\ref{fig:3O_T_Ii_Ij1} and Fig.~\ref{fig:3O_T_Ii_Ij2}  for selected momentums. 
It shows that all the kernel functions are decay with $k \eta$, and the amplitude of  $(I^{(3)}_4)^2$ is the largest.

\section{The energy density spectrum of third order induced gravitational waves}\label{sec:Ph}
In this section, we will study the energy density spectrum of the third order induced gravitational waves. An explicit expression of the power spectrum $\mathcal{P}_{h}^{(3)}(\eta, \mathbf{k})$ is presented. As we shown in Sec.~\ref{sec:eh30}, there are four kinds of source terms for third order induced gravitational waves, the formal expression of $h^{\lambda,(3)}$ is given in Eq.~(\ref{eq:h30}). The two point function $\langle h^{\lambda,(3)} h^{\lambda',(3)}\rangle$ can be expressed as the sum of sixteen terms $\langle h^{\lambda,(3)} h^{\lambda',(3)}\rangle=\sum^{4}_{i,j=1}\langle h^{\lambda,(3)}_ih_j^{\lambda',(3)} \rangle=\langle h_1^{\lambda,(3)} h_1^{\lambda',(3)}\rangle+\langle h_1^{\lambda,(3)} h_2^{\lambda',(3)}\rangle+\cdot \cdot \cdot$. The power spectrum of the third order gravitational waves is given by,
\begin{equation}\label{eq:p}
	\begin{aligned}
		\mathcal{P}_{h}^{(3)}(\eta, \mathbf{k})=\sum_{i,j=1}^{4}\mathcal{P}_{h}^{ij}(\eta, \mathbf{k}).
	\end{aligned}
\end{equation}
It is known that the two-point correlation function and the power spectra are related by $\langle h_*^{\lambda}(\eta,\mathbf{k}) h_*^{\lambda^{\prime}}(\eta,\mathbf{k}')\rangle= \delta^{\lambda \lambda'}\delta\left(\mathbf{k}+\mathbf{k}'\right) \frac{2 \pi^{2}}{k^{3}} \mathcal{P}_{h_*}(\eta, \mathbf{k})$.
Based on the expression of $h^\lambda_i(\eta,\mathbf{k})$ in Eqs.~(\ref{eq:h31})--(\ref{eq:h34}), we obtain the power spectra $\mathcal{P}_{h}^{ij}(\eta, \mathbf{k})$ in the form of
\begin{equation}\label{eq:Ppij}
	\begin{aligned}
		\mathcal{P}_{h}^{ij}(\eta, k)&=\frac{k^3(2\pi^2)^2}{2}\int \frac{\mathrm{d}^{3} p \mathrm{~d}^{3} q\mathrm{d}^{3} p' \mathrm{~d}^{3} q'}{(2 \pi)^{6}}\mathbb{P}^{ij}(\mathbf{k},\mathbf{p},\mathbf{p}',\mathbf{q},\mathbf{q}') \mathcal{C}(\mathbf{k},\mathbf{k}',\mathbf{p},\mathbf{p}',\mathbf{q},\mathbf{q}')\\
		&\times I_i^{(3)}(|\mathbf{k}-\mathbf{p}|,|\mathbf{p}-\mathbf{q}|, p, q,\eta)I_j^{(3)}(|\mathbf{k}'-\mathbf{p}'|,|\mathbf{p}'-\mathbf{q}'|, p', q',\eta)  \ .
	\end{aligned}
\end{equation}
It is found that the $\mathcal{P}_{h}^{ij}(\eta, \mathbf{k})$ for different $i,j$ is the determined by a polynomial $\mathbb{P}^{ij}(\mathbf{k},\mathbf{p},\mathbf{p}',\mathbf{q},\mathbf{q}')$. We obtain the expressions of  $\mathbb{P}^{ij}(\mathbf{k},\mathbf{p},\mathbf{p}',\mathbf{q},\mathbf{q}')$ by making use of the properties of polarization vectors and polarization tensors in Eq.~(\ref{eq:h32}) and Eq.~(\ref{eq:h33}). The $\mathcal{C}(\mathbf{k},\mathbf{k}',\mathbf{p},\mathbf{p}',\mathbf{q},\mathbf{q}')$ is derived from 
the six-point correlation function, and its explicit expression is shown in Eq.~(\ref{eq:C}). 

For illustration, we show the expression of $\mathcal{P}^{ij}(\mathbf{k},\mathbf{p},\mathbf{p}',\mathbf{q},\mathbf{q}')$ for the first term of 	$\mathcal{P}_{h}^{ij}(\eta, k)$ in Eq.~(\ref{eq:C}), 
\begin{equation}\label{eq:Pex} 
	\begin{aligned}
			\mathcal{P}^{i  j}_h & = \frac{k^3}{32 \pi^2} \int \frac{{\rm d}^3 p {\rm d}^3 q}{\left| \mathbf{k} - \mathbf{p} \right|^3 \left| \mathbf{p} - \mathbf{q} \right|^3 q^3} \Big\{ \mathbb{P}^{i  j} \left( k,	\mathbf{p}, - \mathbf{p}, \mathbf{q}, - \mathbf{q} \right) P_{\Phi}     \left( \left| \mathbf{k} - \mathbf{p} \right| \right) P_{\Phi} \left( \left|            \mathbf{p} - \mathbf{q} \right| \right) P_{\Phi} (q)  \\
		& \times I^{(3)}_i \left(	\left| \mathbf{k} - \mathbf{p} \right|, \left| \mathbf{p} - \mathbf{q} 	\right|, p,  q, \eta \right) I^{(3)}_j \left( \left| \mathbf{k} - \mathbf{p}	\right|, \left| \mathbf{p} - \mathbf{q} \right|, p,  q, \eta \right)  \Big\}   \\
		& = \frac{1}{2 \pi} \int_0^{\infty} {\rm d} v \int_0^{\infty}{\rm d} \bar{v} \int^{1 + \bar{v} v}_{| 1 -  \bar{v} v |} {\rm d} w \int^{1 + v}_{| 1 - v |}{\rm d} u \int_{\bar{u}_-}^{\bar{u}_+} {\rm d} \bar{u} \Big\{ \frac{w}{u^2 \bar{u}^2 v \bar{v}^2 \sqrt{Y (1 - X^2)}} \mathbb{P}^{ij} \left( k, \mathbf{p}, - \mathbf{p}, \mathbf{q}, - \mathbf{q} \right)   \\
		&  \times I^{(3)}_i \left( u, v, \bar{u}, \bar{v}, x \right) I^{(3)}_j \left( u, v, \bar{u}, \bar{v}, x \right) P_{\Phi} (k  u) P_{\Phi} (k \bar{u} v) P_{\Phi}(k \bar{v} v) \Big\}~,                        
	\end{aligned}
\end{equation}
where 
\begin{equation}
	\begin{aligned}
		\bar{u}_{\pm} & = \Bigg( 1 + \bar{v}^2 - \frac{(1 + v^2 - u^2) (1 + (\bar{v} v)^2 - w^2)}{2 v^2}   \\
		&  \pm 2 \bar{v} \sqrt{\left( 1 - \left( \frac{1 + v^2 - u^2}{2 v} \right)^2 \right) \left( 1 - \left( \frac{1 + (\bar{v}  v)^2 - w^2}{2 \bar{v} v} \right)^2 \right)}   \Bigg)^{\frac{1}{2}}~,
	\end{aligned}
\end{equation}  

\begin{equation}
	\begin{aligned}
		X & = \left(- 1 + u^2 + v^2 - 2 \bar{u}^2 v^2 +v^2   \bar{v}^2 + u^2 v^2 \bar{v}^2 - v^4\bar{v}^2 + w^2 - u^2 w^2 + v^2 w^2\right)\\
		&\times \left(1 - 2 u^2 + u^4 -   2 v^2 - 2 u^2 v^2 + v^4) (1 - 2 v^2 \bar{v}^2 + v^4 \bar{v}^4 - 2 w^2 - 2 v^2  \bar{v}^2 w^2 + w^4 \right)^{-\frac{1}{2}}~,
	\end{aligned}
\end{equation} 

\begin{equation}
	\begin{aligned}
		Y = (1 - 2 u^2 + u^4 - 2 v^2 - 2 u^2 v^2 + v^4) (1	- 2 v^2 \bar{v}^2 + v^4 \bar{v}^4 - 2 w^2 - 2	v^2 \bar{v}^2 w^2 + w^4)~.
	\end{aligned}
\end{equation} 
We have used the substitution as follows in the first equal,
\begin{equation}
	\begin{aligned}
		\mathbf{k}'=-\mathbf{k} \ , \ \mathbf{p}'=-\mathbf{p} \ , \ \mathbf{q}'=-\mathbf{q} ~,
	\end{aligned}
\end{equation} 
and the substitution as follows in the second equal,
\begin{equation}
	\begin{aligned}
		|\mathbf{k}-\mathbf{p}|=uk \ , |\mathbf{k}-\mathbf{q}|=wk \ , \ |\mathbf{p}-\mathbf{q}|=\bar{u}p=\bar{u}vk   \ , \  
		 q=\bar{v}p=\bar{v}vk~.\label{kpq}
	\end{aligned}
\end{equation}   

\begin{figure}
	\centering
	\includegraphics[scale=0.65]{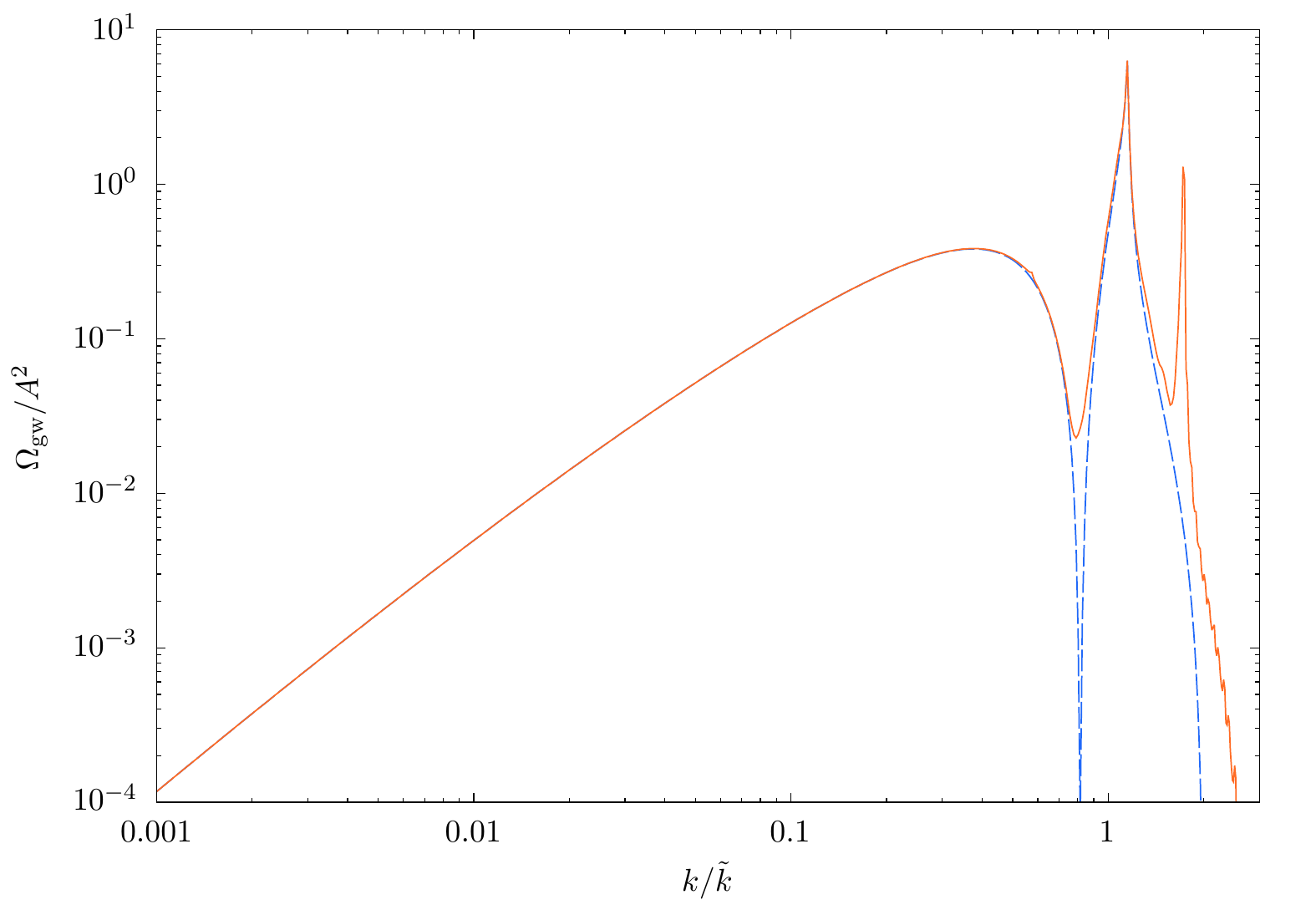}
	\caption{Energy density fractions of induced gravitational waves to the second order (blue curve) and to the third order (orange curve) as function of $\tilde{k}$ for $x=k\eta=1000$. We have set $A=0.001$ in the plot.
	}\label{fig:power_spectrum}
\end{figure}

\begin{figure}
	\centering
	\includegraphics[scale=0.65]{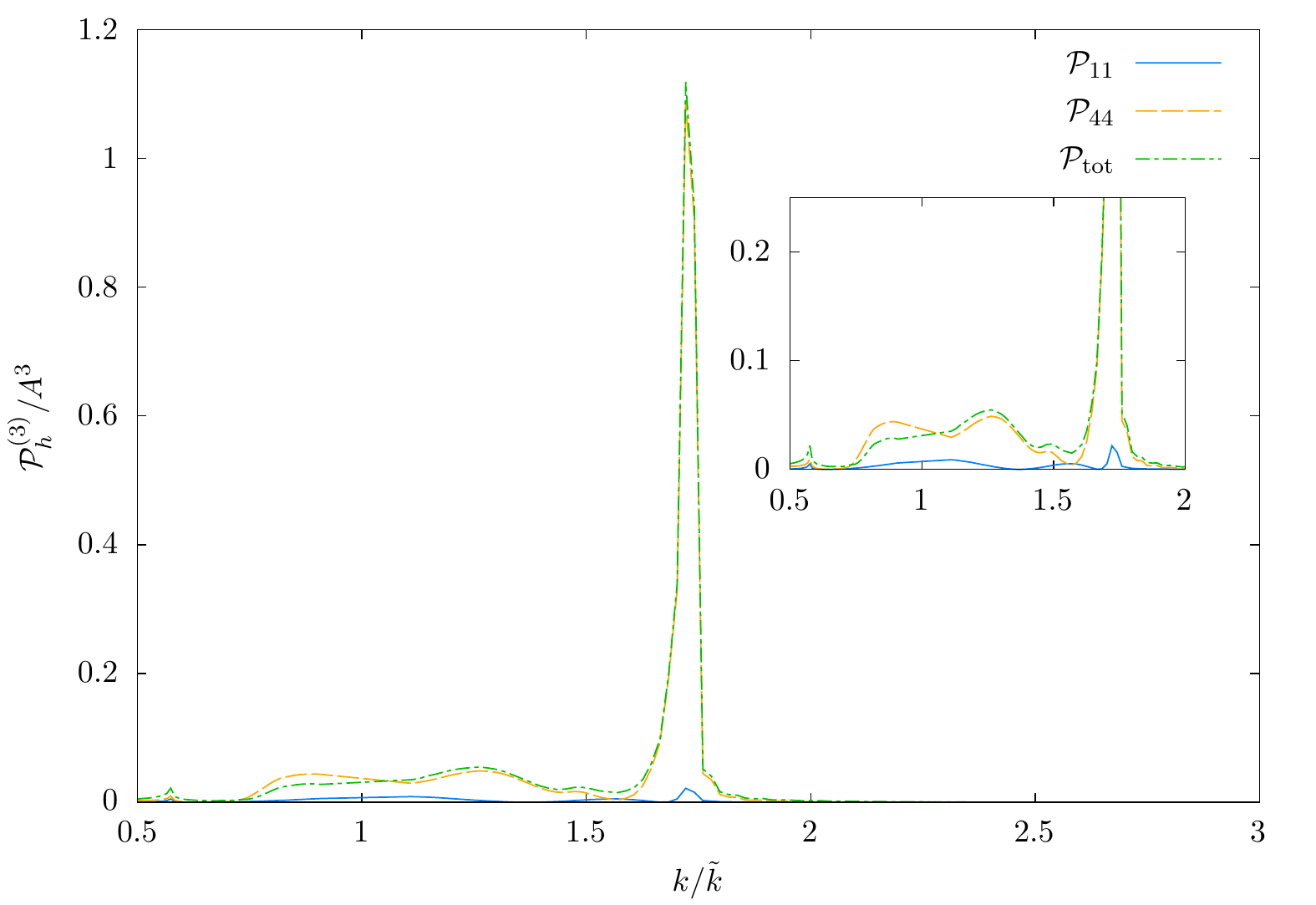}
	\caption{Here, we let $A=0.001$ and $k\eta=1000$. The blue curve and orange curve represent the power spectra sourced by first order scalar perturbation and second order scalar perturbations, respectively. The green curve represents the total power spectrum of third order induced gravitational waves. It shows that the third order gravitational waves sourced by the second order scalar perturbations dominate the two point function $\langle h^{\lambda,(3)}h^{\lambda',(3)} \rangle$ and corresponding energy density spectrum of third order scalar induced gravitational waves}.\label{fig:scalar_dominated}
\end{figure}

In order to obtain a specific result of $\mathcal{P}^{(3)}_{h}(\eta, \mathbf{k})$, we can choose power spectrum of the first order scalar perturbation $	P_{\Phi}(k)$ in the form of a monochromatic power spectrum,
\begin{equation}\label{eq:Pphi}
	\begin{aligned}
		P_{\Phi}(k)=Ak_{*}\delta(k-k_{*}) \ .
	\end{aligned}
\end{equation}
In this case, the power spectrum $\mathcal{P}^{i j}_h$ reduce to 
\begin{eqnarray}
	\mathcal{P}^{i  j}_h(x,\tilde{k}) & = & \frac{\mathcal{A}^3 \tilde{k}^3}{2 \pi}	\Theta (3 - \tilde{k})\int_{\left| 1 - \frac{1}{\tilde{k}}		\right|}^{\min \left\{ \frac{2}{\tilde{k}}, 1 + \frac{1}{\tilde{k}}		\right\}}   {\rm d} v \nonumber \int_{w_-}^{w_+} {\rm d} w\left(\frac{ v w}{\sqrt{Y (1 - X^2)}} \mathbb{P}^{i  j} \left( \mathbf{k}, \mathbf{p}, -	\mathbf{p}, \mathbf{q}, - \mathbf{q} \right) \right. \\
	& &\left. \times \    I^{(3)}_i\left(u,v,\bar{u},\bar{v},x\right)I^{(3)}_j\left(u,v,\bar{u},\bar{v},x\right) \right)_{u=\frac{1}{\tilde{k}}, \tilde{u}=\tilde{v}=\frac{1}{v \tilde{k}}} ~, \label{Z41}
\end{eqnarray}
where we have defined $\tilde{k}\equiv \frac{k}{k_*}$ and 
\begin{equation}
	\begin{aligned}
	w_{\pm} & = \Bigg(\frac{1}{2} + \frac{3}{2 \tilde{k}^2} - \frac{1}{2}
		v^2 \pm \frac{1}{2 v} \sqrt{\left( v^2 -
			\frac{4}{\tilde{k}^2} \right) \left( v^2 - \left( 1 -
			\frac{1}{\tilde{k}} \right)^2 \right) \left( v^2 - \left( 1 +
			\frac{1}{\tilde{k}} \right)^2 \right)}\Bigg)^{\frac{1}{2}}~.
	\end{aligned}
\end{equation}
For illustration, we presented the expression of power spectrum $\mathcal{P}^{ij}_h$ in Eq.~(\ref{Z41}) corresponding with the first term in Eq.~(\ref{eq:C}) only. Note that Eq.~(\ref{eq:C}) comes from the Wick’s theorem of the six point correlation function. The last nine terms in Eq.~(\ref{eq:C}) are proportional to $\delta(k-k_*)$, so we only need to calculate the first six terms. We obtain a complete expression of power spectrum $\mathcal{P}^{ij}_h$ in Eq.~(\ref{Z41}) via substituting  
\begin{equation}\label{Z37}
	\begin{aligned}
	\mathbb{P}^{i  j} \left( \textbf{k}, \textbf{p},  \textbf{p}',	\textbf{q},  \textbf{q}' \right)& \rightarrow \mathbb{P}^{i  j}	\left( \textbf{k}, \textbf{p}, - \textbf{p}, \textbf{q}, - \textbf{q}	\right) +\mathbb{P}^{i  j} \left( \textbf{k}, \textbf{p}, -	\textbf{p}, \textbf{q}, \textbf{q} - \textbf{p} \right) +\mathbb{P}^{i		 j} \left( \textbf{k}, \textbf{p}, \textbf{p} - \textbf{q} -	\textbf{k}, \textbf{q}, - \textbf{q} \right)   \\
	&+\mathbb{P}^{i  j} \left( \textbf{k}, \textbf{p}, \textbf{q} -	\textbf{k}, \textbf{q}, \textbf{q} - \textbf{p} \right) +\mathbb{P}^{i		 j} \left( \textbf{k}, \textbf{p}, \textbf{p} - \textbf{k} -	\textbf{q}, \textbf{q}, \textbf{p} - \textbf{k} \right)   \\
&+\mathbb{P}^{i		 j} \left( \textbf{k}, \textbf{p}, \textbf{q} - \textbf{k},	\textbf{q}, - \textbf{k} + \textbf{p} \right)~,  
	\end{aligned}
\end{equation}
\begin{equation}\label{Ij}
	\begin{aligned}
		I_j^{(3)}(|\mathbf{k}'-\mathbf{p}'|&,|\mathbf{p}'-\mathbf{q}'|, p', q',\eta)  \rightarrow  I_j^{(3)}(|\mathbf{k}-\mathbf{p}|,|\mathbf{p}-\mathbf{q}|,  p, q,\eta)+I_j^{(3)}(|\mathbf{k}-\mathbf{p}|,|\mathbf{p}-\mathbf{q}|,  p, |\mathbf{p}-\mathbf{q}|,\eta) \\
		&+I_j^{(3)}(|\mathbf{k}-\mathbf{p}|,|\mathbf{p}-\mathbf{q}|, |\textbf{p} - \textbf{q} -	\textbf{k}|, q,\eta)+I_j^{(3)}(|\mathbf{k}-\mathbf{p}|,|\mathbf{p}-\mathbf{q}|,  |\mathbf{q}-\mathbf{k}|, |\mathbf{p}-\mathbf{q}|,\eta) \\
		&+I_j^{(3)}(|\mathbf{k}-\mathbf{p}|,|\mathbf{p}-\mathbf{q}|, |\textbf{p} - \textbf{q} -	\textbf{k}|, |\mathbf{p}-\mathbf{k}|,\eta) \\
		&+I_j^{(3)}(|\mathbf{k}-\mathbf{p}|,|\mathbf{p}-\mathbf{q}|, |\textbf{q} - 	\textbf{k}|, |\mathbf{p}-\mathbf{k}|,\eta) ~. 
	\end{aligned}
\end{equation}
In the case of monochromatic power spectrum $	P_{\Phi}(k)=Ak_{*}\delta(k-k_{*})$, Eq.~(\ref{Z37}) can be expressed as 
\begin{equation}\label{Z3}
	\begin{aligned}
		\mathbb{P}^{i  j} \left( \textbf{k}, \textbf{p},  \textbf{p}',	\textbf{q},  \textbf{q}' \right) \rightarrow \mathbb{P}^{i  j}_{a}=\left(\mathbb{P}^{i  j}_{1} \ , \ \mathbb{P}^{i  j}_{2} \ , \ \mathbb{P}^{i  j}_{3} \right) ~,  
	\end{aligned}
\end{equation}
where the explicit expressions of $\mathbb{P}^{ij}_{a},(a=1,2,3)$ in Eq.~(\ref{Z3}) are shown in the Appendix~\ref{sec:D}. The Eq.~(\ref{Ij}) can be expressed as 
\begin{equation}
	\begin{aligned}
		I_j^{(3)}&(|\mathbf{k}'-\mathbf{p}'|,|\mathbf{p}'-\mathbf{q}'|, p', q',\eta)  \rightarrow I^{(3),a}_j\left( |\mathbf{k}'-\mathbf{p}'|,|\mathbf{p}'-\mathbf{q}'|, p', q',\eta \right) \\
		&=2\left(I_j^{(3),1}(k_*,k_*,  p, k_*,\eta),I_j^{(3),2}(k_*,k_*,  |\mathbf{k}-\mathbf{q}|, k_*,\eta),I_j^{(3),3}(k_*,k_*, |\textbf{p} - \textbf{q} -	\textbf{k}|, k_*,\eta) \right) ~.
	\end{aligned}
\end{equation}
Namely,
\begin{equation}\label{Ij2}
	\begin{aligned}
		&I^{(3),1}_j\left( u',v',\bar{u}',\bar{v}',\eta \right) = 2I_j^{(3)}\left(\frac{1}{\tilde{k}},v,\frac{1}{v\tilde{k}},\frac{1}{v\tilde{k}},\eta\right)~, 
		\\
		& I^{(3),2}_j\left( u',v',\bar{u}',\bar{v}',\eta \right) =2I_j^{(3)}\left(\frac{1}{\tilde{k}},w,\frac{1}{w\tilde{k}},\frac{1}{w\tilde{k}},\eta\right)~,  \\
		&I^{(3),3}_j\left( u',v',\bar{u}',\bar{v}',\eta \right)= 2I_j^{(3)}\left(\frac{1}{\tilde{k}}, \left(1-v^2-w^2+3/\tilde{k}^2\right)^{\frac{1}{2}},\frac{1}{\tilde{k}}\left(1-v^2-w^2+3/\tilde{k}^2\right)^{-\frac{1}{2}} , \right.  \\
		&\left. \frac{1}{\tilde{k}}\left(1-v^2-w^2+3/\tilde{k}^2\right)^{-\frac{1}{2}},\eta\right)  ~,
	\end{aligned}
\end{equation}
where we have defined 
\begin{equation}
	\begin{aligned}
		|\mathbf{k}'-\mathbf{p}'|=u'k' \ ,  \  |\mathbf{p}'-\mathbf{q}'|=\bar{u}'p'=\bar{u}'v'k'   \ , \   q'=\bar{v}'p'=\bar{v}'v'k'~.
	\end{aligned}
\end{equation}   
Substituting Eq.~(\ref{Z3}) and Eq.~(\ref{Ij2}) into Eq.~(\ref{Z41}), we obtain the complete expression of power spectrum $\mathcal{P}^{ij}_h$ for monochromatic primordial power spectrum
\begin{eqnarray}
	\mathcal{P}^{i  j}_h(x,\tilde{k}) & = & \frac{\mathcal{A}^3 \tilde{k}^3}{2 \pi}	\Theta (3 - \tilde{k})\int_{\left| 1 - \frac{1}{\tilde{k}}		\right|}^{\min \left\{ \frac{2}{\tilde{k}}, 1 + \frac{1}{\tilde{k}}		\right\}}   {\rm d} v \nonumber \int_{w_-}^{w_+} {\rm d} w\left(\frac{ v w}{\sqrt{Y (1 - X^2)}}  \right. \\
	& &\left. \times \ I^{(3)}_i\left(u,v,\bar{u},\bar{v},x\right) \ \sum^{3}_{a=1} \mathbb{P}^{i  j}_{a} \  I^{(3),a}_j\left( u',v',\bar{u}',\bar{v}',\eta \right) \ \right)_{u=\frac{1}{\tilde{k}}, \bar{u}=\bar{v}=\frac{1}{v \tilde{k}}} ~. \label{Z46}
\end{eqnarray}
The Eq.~(\ref{Z46}) is composed of three parts. First, the integrals and integral measures, come from the four three-dimensional momentum integrals in Eq.~(\ref{eq:Ppij}). As shown in Eq.~(\ref{eq:Pex}), the integrals of $\mathbf{p}'$ and $\mathbf{q}'$ can be evaluated in terms of the three-dimensional delta functions in Eq.~(\ref{eq:C}). Three one-dimensional integrals in the second equal of Eq.~(\ref{eq:Pex}) can be evaluated in terms of the monochromatic power spectrum. Therefore, there are only two one-dimensional integrals in Eq.~(\ref{Z46}). Second, the polynomials $\mathbb{P}^{i  j}_{a}$, come from the decomposed operators in Eq.~(\ref{eq:h31})$\sim$ Eq.~(\ref{eq:h34}). As we mentioned before, the summation of index $a$ comes from the wick theorem in Eq.~(\ref{eq:C}). The third part is the kernel functions $I^{(3)}_i\left(u,v,\bar{u},\bar{v},x\right)$ and $I^{(3),a}_j\left( u',v',\bar{u}',\bar{v}',\eta \right)$. The kernel functions $I^{(3)}_i\left(u,v,\bar{u},\bar{v},x\right)$ come from the kernel functions $I_i^{(3)}(|\mathbf{k}-\mathbf{p}|,|\mathbf{p}-\mathbf{q}|, p, q,\eta)$ in Eq.~(\ref{eq:Ppij}), they are the functions of $\mathbf{k}$, $\mathbf{p}$, and $\mathbf{q}$. When we substitute Eq.~(\ref{eq:C}) into Eq.~(\ref{eq:Ppij}), the kernel functions $I_i^{(3)}(|\mathbf{k}-\mathbf{p}|,|\mathbf{p}-\mathbf{q}|, p, q,\eta)$ won't change. On the contrary, the kernel functions $I_j^{(3)}(|\mathbf{k}'-\mathbf{p}'|,|\mathbf{p}'-\mathbf{q}'|, p', q',\eta)$ are the functions of $\mathbf{k}'$, $\mathbf{p}'$, and $\mathbf{q}'$, they will change to Eq.~(\ref{Ij}) when we substitute Eq.~(\ref{eq:C}) into Eq.~(\ref{eq:Ppij}). In the case of monochromatic power spectrum, $I_j^{(3)}(|\mathbf{k}'-\mathbf{p}'|,|\mathbf{p}'-\mathbf{q}'|, p', q',\eta)$ can be simplified to Eq.~(\ref{Ij2}). In the end of this section, we calculate the third order power spectrum $\mathcal{P}^{(3)}_h$, i.e., 
\begin{equation}
	\mathcal{P}^{(3)}_h = \sum_{i,j=1}^{4}\mathcal{P}^{ij}_h(x,\tilde{k})~.\label{Z38} 
\end{equation}
The fraction of the gravitational waves energy density per logarithmic wavelength is given by 
\begin{equation}\label{eq:Omega}
	\begin{aligned}
		\Omega_{\mathrm{GW}}(\eta, k)=\frac{\rho_{\mathrm{GW}}(\eta, k)}{\rho_{\mathrm{tot}}(\eta)}=\frac{1}{24}\left(\frac{k}{a(\eta) H(\eta)}\right)^{2} {\mathcal{P}_{h}(\eta, k)} \ ,
	\end{aligned}
\end{equation}
where
\begin{equation}
	\begin{aligned}
		{\mathcal{P}_{h}(\eta, k)}=\frac{1}{4}{\mathcal{P}^{(2)}_{h}(\eta, k)}+\frac{1}{36}{\mathcal{P}^{(3)}_{h}(\eta, k)}+O \left({\mathcal{P}^{(4)}_{h}}\right)\ .
	\end{aligned}
\end{equation}
Evaluating power spectrum ${\mathcal{P}_{h}(\eta, k)}$ in Eq.~(\ref{eq:Omega}) by using Eqs.~(\ref{Z41}) and (\ref{Z38}), we obtain the energy density fraction of the third order induced gravitational waves. As it is shown in Fig.~\ref{fig:power_spectrum}, the shape of density fraction is  different when considering the third order gravitational waves. In Fig.~\ref{fig:scalar_dominated}, we present the third order power spectra of selected sources. It shows that the third order gravitational waves sourced by the second order scalar perturbations dominate the two point function $\langle h^{\lambda,(3)}h^{\lambda',(3)} \rangle$ and corresponding energy density spectrum of third order scalar induced gravitational waves.


\section{Conclusions and discussions}\label{sec:CaD}
In this paper, we studied the third order scalar induced gravitational waves. The source terms of the third order scalar induced  gravitational waves were completely considered. For illustration, we presented energy density spectrum of the third order gravitational waves for a monochromatic primordial power spectrum. The explicit integral expressions given in Eqs.~(\ref{eq:Pex}) and (\ref{Z41}) are shown to be useful for evaluating the energy density spectrum of the third order gravitational waves. Our results are different from  the pioneer' study \cite{Yuan:2021qgz} in principle, which neglected all the second cosmological perturbations. It was found that the third order gravitational waves sourced by the second order scalar perturbations dominates the energy density spectrum.
The source term of the third order induced gravitational waves $h^{(3)}_{ij}$ is completely different from the second order induced gravitational waves $h^{(2)}_{ij}$.  Because the first order scalar, vector, and tensor perturbations are independent, it is nothing wrong to set the vector and tensor perturbations to be zero. The second order induced gravitational wave sourced by the first order scalar perturbations only is well-defined. However, for the third order gravitational wave $h^{(3)}_{ij}$, the sources include the first order scalar perturbations and three types of the second order perturbations. The second order perturbations induced by the first order scalar perturbations are not independent at all. One can not set any the second order perturbation to be zero. For instance, if we set the second order vector perturbation $V^{(2)}_{j}$  to zero, the right hand side of equation of motion will be zero. It will bring an unphysical constraint to the first order scalar perturbation $\phi^{(1)}$.  This discussion can also be applied to higher order gravitational waves.

For second order scalar induced gravitational waves, one can derive the explicit analytical expression of $I^{(2)}$\cite{Kohri:2018awv}. For the third order scalar induced gravitational waves, it is difficult to obtain the explicit analytical expressions for the third order kernel functions $I^{(3)}_i$. However, we can still obtain the asymptotic properties of the third order kernel functions $I^{(3)}_i$ in terms of the numerical result. We found that the influence of different integral upper limit $x$ $(x\sim 1000)$ in Eqs.~(\ref{eq:kerneli}) is negligible for the peaks in Fig.~\ref{fig:power_spectrum} and Fig.~\ref{fig:scalar_dominated}. It may suggest that $\Omega^{(3)}_{GW}$ is independent of $x$. The third order kernel functions $I^{(3)}_i$ decay as $\frac{1}{x}$ for large $x$.
A semianalytic calculation of third order scalar induced gravitational waves might be presented in the future.

We calculated four types of kernel functions of the third order induced gravitational waves, there are sixteen combinations of the product of kernel functions $I^{(3)}_iI^{(3)}_j$. As shown in Fig.~\ref{fig:3O_T_Ii_Ij1} and Fig.~\ref{fig:3O_T_Ii_Ij2}, the $(I^{(3)}_3)^2$ is much larger than second order kernel function $(I^{(2)}_h)^2$. Therefore, it is expected that the second order vector perturbations would influence the amplitude of the third order gravitational waves.
However, for a monochromatic primordial power spectrum, we found that the $\mathbb{P}^{33}_h$ is zero. In this case, the corresponding power spectrum $\mathcal{P}^{33}_h$ has no relevance with the  power spectrum of the third order scalar induced gravitational waves. In this sense, it is necessary to consider other primordial power spectrum in further study, for example, the power-law spectrum. And non-trivial results might be obtained for a general primordial power spectrum. 

We considered the third order gravitational waves induced by the first order scalar perturbations. In principle, there should be second order primordial curvature perturbation \cite{Inomata:2020cck} that could also induce the third order gravitational waves. Since the correlation function between the first and second order scalar perturbations refers to a carefull calculation in a inflation model, relevant studies might be given in the future. 

As we mentioned in Sec.~\ref{sec:Ph}, the last nine terms in Eq.~(\ref{eq:C}) are proportional to $\delta(k-k_*)$, so we only need to calculate the first six terms. Here, we give an explicit explanation as follow. For third order induced gravitational waves, the formal expression of power spectrum is given in Eq.~(\ref{eq:Ppij}). In Eq.~(\ref{eq:Ppij}), $\mathcal{C}(\mathbf{k},\mathbf{k}',\mathbf{p},\mathbf{p}',\mathbf{q},\mathbf{q}')$ is derived from the six point function $\langle\Phi_{\mathbf{k}-\mathbf{p}} \Phi_{\mathbf{p}-\mathbf{q}} \Phi_{\mathbf{q}}\Phi_{\mathbf{k}'-\mathbf{p}'} \Phi_{\mathbf{p}'-\mathbf{q}'} \Phi_{\mathbf{q}'}\rangle$, and its explicit expression is shown in Eq.~(\ref{eq:C}). For example, the first term in Wick's expansion Eq.~(\ref{eq:C}) is
\begin{equation}\label{2}
	\left( \frac{1}{(k-p)^3(p-q)^3q^3}\delta(\mathbf{p}+\mathbf{p}')\delta(\mathbf{q}+\mathbf{q}')P_{\Phi}(k-p)P_{\Phi}(p-q)P_{\Phi}(q)\right) \ .
\end{equation}
It contains two three dimensional delta functions $\delta(\mathbf{p}+\mathbf{p}')$ and $\delta(\mathbf{q}+\mathbf{q}')$. Substitute the Eq.~(\ref{2}) into Eq.~(\ref{eq:Ppij}), we can integral out two three dimensional delta functions and obtain:
\begin{equation}\label{3}
	\mathbf{p}'=-\mathbf{p}  \ , \ \mathbf{q}'=-\mathbf{q} \ .
\end{equation}
As shown in Eq.~(\ref{Z37}) $\sim$ Eq.~(\ref{Z46}), the first six terms in Wick's expansion in Eq.~(\ref{eq:C}) can be study in the say way.
But for the last nine terms in Wick's expansion in Eq.~(\ref{eq:C}), we will encounter $\delta(k-k_*)$. For example, the last term in Wick's expansion in Eq.~(\ref{eq:C}) is
\begin{equation}\label{4}
	\left( \frac{1}{(k-p)^3(p-q)^3(-k-p')^3}\delta(\mathbf{p})\delta(-\mathbf{k}-\mathbf{q}')P_{\Phi}(k-p)P_{\Phi}(p-q)P_{\Phi}(-k-p')\right) \ .
\end{equation}
It also contains two three dimensional delta functions $\delta(\mathbf{p})$ and $\delta(-\mathbf{k}-\mathbf{q}')$. Substituting the Eq.~(\ref{4}) into Eq.~(\ref{eq:Ppij}) and integrated out two three dimensional delta functions, we obtain
\begin{equation}\label{5}
	\mathbf{p}=0  \ , \ \mathbf{q}'=-\mathbf{k} \ .
\end{equation}
Then the first primordial power spectrum $P_{\Phi}(k-p)$ in Eq.~(\ref{4}) will become $P_{\Phi}(k)=Ak_*\delta(k-k_*)$. This is the origin of the delta function $\delta(k-k_*)$ in the power spectrum. The delta function $\delta(k-k_*)$ apears in the last nine terms in Eq.~(\ref{eq:C}), and it is the direct result of the Wick's theorem. However, we don't need to worry about these nine terms, they correspond to the bubble diagrams, namely, we will encounter the integrals such as $\int d^3p \mathcal{P}_{\Phi}(p)$ in these nine terms. Here, we have neglected these bubble diagrams and corresponding unphysical delta functions in our manuscript.

It has to be clarified that this paper studied the power spectrum of the third order gravitational waves instead of the third order power spectrum of gravitational waves. More precisely, the power spectrum of the third order gravitational waves comes from the two point function $\langle h^{\lambda,(3)}h^{\lambda',(3)} \rangle$ of third order induced gravitational waves. We have studied it systematically in this paper, it only include the contribution from the third order gravitational waves $h^{\lambda,(3)}$. However, the third order power spectrum of gravitational waves is composed of $\langle h^{\lambda,(3)}h^{\lambda',(3)} \rangle$ and $\langle h^{\lambda,(2)}h^{\lambda',(4)} \rangle$.
An incomplete study of the third order power spectrum of gravitational waves was given in Refs.~\cite{Yuan:2019udt,Yuan:2021qgz}, which neglected  all the contributions from the higher order cosmological perturbations. Perhaps, a complete study on the third order power spectrum of induced gravitational waves might be presented in the future.

\appendix
\section{Decomposition Operators}\label{sec:A}
For an arbitrary spatial tensor filed $S_{ij}$ on FRW spacetime, it could be decomposed as scalar, vector and tensor modes
\begin{equation}
	S_{i j}=S_{i j}^{(H)}+2 \delta_{i j} S^{(\Psi)}+2 \partial_{i} \partial_{j} S^{(E)}+\partial_{j} S_{i}^{(C)}+\partial_{i} S_{j}^{(C)} \ .
\end{equation}
We define the decomposed operator to fulfill this decomposition.
\begin{equation}
	S_{i j}^{(H)} \equiv\Lambda_{ij}^{kl}S_{k l}=\left(\mathcal{T}_{i}^{k} \mathcal{T}_{j}^{l}-\frac{1}{2} \mathcal{T}_{i j} \mathcal{T}^{k l}\right) S_{k l} \ \ , \ \ S^{(\Psi)} \equiv \frac{1}{4} \mathcal{T}^{k l} S_{k l} \ .
\end{equation}
\begin{equation}
	S^{(E)} \equiv \frac{1}{2} \Delta^{-1}\left(\partial^{k} \Delta^{-1} \partial^{l}-\frac{1}{2} \mathcal{T}^{k l}\right) S_{l k} \ \ , \ \ S_{i}^{(C)} \equiv \Delta^{-1} \partial^{l} \mathcal{T}_{i}^{k} S_{l k} \ .
\end{equation}
The transverse operators is defined by
\begin{equation}
	\mathcal{T}_{j}^{i} \equiv \delta_{i}^{i}-\partial^{i} \Delta^{-1} \partial_{j} \ .
\end{equation} 
This type of decomposed operator can only use in cosmology. For arbitrary spacetime, we can't distinguish between $S_{i}^{(C)}$ and $S^{(E)}$.

It is not difficult to calculate the decomposed operator in momentum space. For tensor mode, the corresponding decomposed operator can be expressed as
\begin{equation}
	\begin{aligned}
		\Lambda_{i j}^{l m}&=\mathcal{T}_{i}^{l} \mathcal{T}_{j}^{m}-\frac{1}{2} \mathcal{T}_{i j} \mathcal{T}^{l m} \\
		&=\left(\delta_{i}^{l}-\partial^{l} \Delta^{-1} \partial_{i}\right)\left(\delta_{j}^{m}-\partial^{m} \Delta^{-1} \partial_{j}\right)-\frac{1}{2}\left(\delta_{i j}-\partial_{i} \Delta^{-1} \partial_{j}\right)\left(\delta^{l m}-\partial^{l} \Delta^{-1} \partial^{m}\right) \ .
	\end{aligned}
\end{equation}
We set $\partial_{i} \rightarrow i k_{i}$ in momentum space
\begin{equation}\label{eq:lamp}
	\begin{aligned}
		\Lambda_{i j}^{l m}(\mathbf{k})&=\left(\delta_{i}^{l}-\frac{k^{l} k_{i}}{|\mathbf{k}|^{2}}\right)\left(\delta_{j}^{m}-\frac{k^{m} k_{j}}{|\mathbf{k}|^{2}}\right)-\frac{1}{2}\left(\delta_{i j}-\frac{k_{i} k_{j}}{|\mathbf{k}|^{2}}\right)\left(\delta^{l m}-\frac{k^{l} k^{m}}{|\mathbf{k}|^{2}}\right) \\
		&=\left(\delta_{i}^{l}-n^{l}(\mathbf{k}) n_{i}(\mathbf{k})\right)\left(\delta_{j}^{m}-n^{m}(\mathbf{k}) n_{j}(\mathbf{k})\right)-\frac{1}{2}\left(\delta_{i j}-n_{i}(\mathbf{k}) n_{j}(\mathbf{k})\right)\left(\delta^{l m}-n^{l}(\mathbf{k}) n^{m}(\mathbf{k})\right) \ ,
	\end{aligned}
\end{equation}
where $n^{i}(\mathbf{k})=k^{i} /|\mathbf{k}|$ is the unit vector with respect to $\mathbf{k}$. In three dimensional momentum space, we can choose a normalized bases $\{n_i(\mathbf{k}),e_i(\mathbf{k}),\bar{e}_i(\mathbf{k}) \}$. They satisfy the following two conditions
\begin{equation}
	\begin{aligned}
		e_{i}(\mathbf{k}) e^{i}(\mathbf{k})=\bar{e}_{i}(\mathbf{k}) \bar{e}^{i}(\mathbf{k})=n_{i}(\mathbf{k}) n^{i}(\mathbf{k})=1 \ , \\
		e_{i}(\mathbf{k}) \bar{e}^{i}(\mathbf{k})=e_{i}(\mathbf{k}) n^{i}(\mathbf{k})=\bar{e}_{i}(\mathbf{k}) n^{i}(\mathbf{k})=0 \ .
	\end{aligned}
\end{equation}
We can express the Kronecker delta in momentum space in terms of the normalized bases
\begin{equation}\label{eq:en}
	\begin{aligned}
		\delta_{i}^{j}=e_{i}(\mathbf{k}) e^{j}(\mathbf{k})+\bar{e}_{i}(\mathbf{k}) \bar{e}^{j}(\mathbf{k})+n_{i}(\mathbf{k}) n^{j}(\mathbf{k}) \ .
	\end{aligned}
\end{equation}
For vector mode, the decomposed operator is given by
\begin{equation}
	\begin{aligned}
		\Delta^{-1} \mathcal{T}_{l}^{r} \partial^{s}= \Delta^{-1} \left(\delta_{l}^{r}-\partial^{r} \Delta^{-1} \partial_{l}\right)\partial^{s} \ .
	\end{aligned}
\end{equation}
In momentum space, we obtain
\begin{equation}\label{eq:Vec}
	\begin{aligned}
		-\frac{ip^s}{|\mathbf{p}|^{2}}\left(\delta_{l}^{r}-\frac{p^rp_l}{|\mathbf{p}|^{2}}\right)=-\frac{in^s(\mathbf{p})}{|\mathbf{p}|}\left(e_{l}(\mathbf{p}) e^{r}(\mathbf{p})+\bar{e}_{l}(\mathbf{p}) \bar{e}^{r}(\mathbf{p})\right) \ .
	\end{aligned}
\end{equation}
For scalar mode, the corresponding decomposed operators are
\begin{equation}
	\begin{aligned}
		-2 \Delta^{-1}\left(\partial^{r} \Delta^{-1} \partial^{s}-\frac{1}{2} \mathcal{T}^{rs}\right)=-2\Delta^{-1}\left(\frac{3}{2}\partial^{r} \Delta^{-1} \partial^{s}-\frac{1}{2} \delta^{rs}\right) \ ,
	\end{aligned}
\end{equation}
\begin{equation}
	\begin{aligned}
		-\frac{1}{2}\mathcal{T}^{rs}=-\frac{1}{2}\left(\delta^{rs}-\partial^{r} \Delta^{-1} \partial^{s}\right) \ .
	\end{aligned}
\end{equation}
In momentum space, we obtain
\begin{equation}
	\begin{aligned}
		\frac{2}{|\mathbf{p}|^2}\left(\frac{3}{2}\frac{p^rp^s}{|\mathbf{p}|^2}-\frac{1}{2} \delta^{rs}\right) \ , \
		-\frac{1}{2}\left(\delta^{rs}-\frac{p^rp^s}{|\mathbf{p}|^2}\right) \ .
	\end{aligned}
\end{equation} 
\section{The second order perturbations}\label{sec:Lower}
In this appendix, we give the explicit expressions of equations of motion and corresponding kernel functions of three kinds of second order perturbations. We use the \texttt{xPand} package \cite{Pitrou:2013hga} to obtain the equations of motion of second order perturbations. 
\subsection{Second order tensor perturbation}
For second order tensor perturbations, the equation of motion is 
\begin{equation}\label{eq:eh2x}
	h_{lm}^{(2)''}(\eta,\mathbf{x})+2 \mathcal{H}  h_{lm}^{(2)'}(\eta,\mathbf{x})-\Delta h_{lm}^{(2)}(\eta,\mathbf{x})=-4 \Lambda_{lm}^{rs} \mathcal{S}^{(2)}_{rs}(\eta,\mathbf{x}) \ ,
\end{equation}
where the source term $\Lambda_{lm}^{rs} \mathcal{S}^{(2)}_{rs}(\eta,\mathbf{x})$ is 
\begin{equation}
	\Lambda_{lm}^{rs} \mathcal{S}^{(2)}_{rs}(\eta,\mathbf{x})=\Lambda_{l m}^{rs}\left(3 \phi^{(1)} \partial_{r} \partial_{s} \phi^{(1)}+\frac{2}{\mathcal{H}}  \phi^{(1)'} \partial_{r} \partial_{s} \phi^{(1)}+\frac{1}{\mathcal{H}^{2}}  \phi^{(1)'} \partial_{r} \partial_{s}  \phi^{(1)'}\right)  \ .
\end{equation}
The corresponding kernel function is 
\begin{equation}
	I_{h}^{(2)}(\bar{u},\bar{v},y)=\frac{4}{p^{2}} \int_{0}^{y} \mathrm{d}\bar{y} \left(\frac{\bar{y}}{y} \sin(y-\bar{y}) f^{(2)}_h(\bar{u},\bar{v},y)\right) \ ,
\end{equation}
where
\begin{equation}
	\begin{aligned}
		f^{(2)}_h(\bar{u},\bar{v},y)=2 T_{\phi}(\bar{u}y) T_{\phi}(\bar{v}y)+\left(\bar{u}y \frac{d}{d(\bar{u}y)}T_{\phi}(\bar{u}y)+T_{\phi}(\bar{u}y)\right)\left(\bar{v}y \frac{d}{d(\bar{v}y)}T_{\phi}(\bar{v}y)+T_{\phi}(\bar{v}y)\right) \ .
	\end{aligned}
\end{equation}
\subsection{Second order vector perturbation}
The equation of motion of second order vector perturbation is 
\begin{equation}\label{eq:eV}
	\begin{aligned}
		V_{l}^{(2)'}(\eta,\mathbf{x})+2 \mathcal{H} V_{l}^{(2)}(\eta,\mathbf{x}) =4 \Delta^{-1} \mathcal{T}_{l}^{r} \partial^{s} \mathcal{S}^{(2)}_{rs}(\eta,\mathbf{x}) \ ,
	\end{aligned}
\end{equation}
where the source term is given by
\begin{equation}
	\begin{aligned}
		\Delta^{-1} \mathcal{T}_{l}^{r} \partial^{s} \mathcal{S}^{(2)}_{rs}(\eta,\mathbf{x})=& \Delta^{-1} \mathcal{T}_{l}^{r} \partial^{s}\Bigg(\partial_{r} \phi^{(1)} \partial_{s} \phi^{(1)}-\frac{1}{ \mathcal{H}}\left(\partial_{r} \phi^{(1)'} \partial_{s} \phi^{(1)}+\partial_{r} \phi^{(1)}  \partial_{s} \phi^{(1)'}\right)\\
		&+4 \phi^{(1)} \partial_{r} \partial_{s} \phi^{(1)}-\frac{1}{ \mathcal{H}^{2}} \partial_{r} \phi^{(1)'} \partial_{s}  \phi^{(1)'}\Bigg) \ .
	\end{aligned}
\end{equation}
After making use of the Fourier transformation, we obtain the equation of motion of second order vector perturbation in momentum space
\begin{equation}\label{eq:eV2}
	\begin{aligned}
		V^{\lambda,(2)'}(\eta,\mathbf{p})+2 \mathcal{H} V^{\lambda,(2)}(\eta,\mathbf{p}) =4i \mathcal{S}_V^{\lambda,(2)}(\eta,\mathbf{p}) \ ,
	\end{aligned}
\end{equation}
where $V^{\lambda,(2)}(\eta,\mathbf{p})=e^{\lambda,l}(\mathbf{p})V_{l}^{(2)}(\eta,\mathbf{p})$, $e^{\lambda,l}(\mathbf{p})$ is the polarization vectors with respect to $\mathbf{p}$. The source term can be written as
\begin{equation}\label{eq:SV}
	\begin{aligned}
		\mathcal{S}_V^{\lambda,(2)}(\eta,\mathbf{p})&=\frac{p^se^{\lambda,r}(\mathbf{p})}{p^2} \mathcal{S}_{rs}(\eta,\mathbf{p})=\int\frac{d^3q}{(2\pi)^{3/2}}\frac{p^se^{\lambda,r}(\mathbf{p})}{p^2}q_rq_s\Bigg(  3\phi^{(1)}(\mathbf{p}-\mathbf{q})  \phi^{(1)}(\mathbf{q})\\
		&+\frac{2}{ \mathcal{H}} \phi^{(1)'}(\mathbf{p}-\mathbf{q})  \phi^{(1)}(\mathbf{q})+\frac{1}{ \mathcal{H}^{2}}\phi^{(1)'}(\mathbf{p}-\mathbf{q})   \phi^{(1)'}(\mathbf{q})
		\Bigg) \ .
	\end{aligned}
\end{equation}
The corresponding kernel function is 
\begin{equation}\label{eq:IV}
	\begin{aligned}
		I_{V}^{(2)}(\bar{u},\bar{v},y)=\frac{4}{p}\int_0^{y} f_V^{(2)}(\bar{u},\bar{v},y)\frac{\bar{y}^2}{y^2} d\bar{y} \ ,
	\end{aligned}
\end{equation}
where
\begin{equation}\label{eq:fV2}
	\begin{aligned}
		f_V(\bar{u},\bar{v},y)= 3T_{\phi}(\bar{u}y)  T_{\phi}(\bar{v}y)+2\bar{u}y \frac{d}{d(\bar{u}y)}T_{\phi}(\bar{u}y)  T_{\phi}(\bar{v}y)+\bar{u}\bar{v}y^2\frac{d}{d(\bar{u}y)}T_{\phi}(\bar{u}y)  \frac{d}{d(\bar{v}y)}T_{\phi}(\bar{v}y) \ .
	\end{aligned}
\end{equation}
\subsection{Second order scalar perturbations}
The equations of motion of second order scalar perturbations are
\begin{equation}
	\begin{aligned}
		\psi^{(2)}(\eta,\mathbf{x})-\phi^{(2)}(\eta,\mathbf{x})=-2 \Delta^{-1}\left(\partial^{r} \Delta^{-1} \partial^{s}-\frac{1}{2} \mathcal{T}^{rs}\right) \mathcal{S}^{(2)}_{rs}(\eta,\mathbf{x}) \ ,
	\end{aligned}
\end{equation}
\begin{equation}
	\begin{aligned}
		\psi^{(2)''}(\eta,\mathbf{x})+3\mathcal{H}  \psi^{(2)'}(\eta,\mathbf{x})-\frac{5}{6} \Delta \psi^{(2)}(\eta,\mathbf{x})+\mathcal{H}  \phi^{(2)'}(\eta,\mathbf{x})+\frac{1}{2} \Delta \phi^{(2)}(\eta,\mathbf{x})\\
		=-\frac{1}{2} \mathcal{T}^{rs} \mathcal{S}^{(2)}_{rs}(\eta,\mathbf{x}) \ ,
	\end{aligned}
\end{equation}
where the source terms are
\begin{equation}
	\begin{aligned}
		&-2 \Delta^{-1}\left(\partial^{r} \Delta^{-1} \partial^{s}-\frac{1}{2} \mathcal{T}^{rs}\right) \mathcal{S}^{(2)}_{rs}(\eta,\mathbf{x})=-2 \Delta^{-1}\left(\partial^{r} \Delta^{-1} \partial^{s}-\frac{1}{2} \mathcal{T}^{rs}\right)\Bigg( \partial_{r} \phi^{(1)} \partial_{s} \phi^{(1)}\\
		&+4 \phi^{(1)} \partial_{r} \partial_{s} \phi^{(1)}
		-\frac{1}{ \mathcal{H}}\left(\partial_{r} \phi^{(1)'} \partial_{s} \phi^{(1)}+\partial_{r} \phi^{(1)} \partial_{s} \phi^{(1)'}\right)-\frac{1}{ \mathcal{H}^{2}} \partial_{r}  \phi^{(1)'} \partial_{s}  \phi^{(1)'}\Bigg)  \ ,
	\end{aligned}
\end{equation}
\begin{equation}
	\begin{aligned}
		-\frac{1}{2} \mathcal{T}^{rs} \mathcal{S}^{(2)}_{rs}(\eta,\mathbf{x})&= \Bigg(\frac{11}{3} \partial_{k} \phi^{(1)} \partial^{k} \phi^{(1)}+24 \mathcal{H} \phi^{(1)}  \phi^{(1)'}+\frac{16}{3} \phi^{(1)} \Delta \phi^{(1)}-\frac{2}{3 \mathcal{H}} \partial_{k}  \phi^{(1)'} \partial^{k} \phi^{(1)}\\
		&+2\left( \phi^{(1)'}\right)^{2}+4 \phi^{(1)} \phi^{(1)''} -\frac{1}{3\mathcal{H}^{2}} \partial_{k} \phi^{(1)'} \partial^{k} \phi^{(1)'}\Bigg)  \\
		&-\frac{1}{2} \mathcal{T}^{rs}\left(3 \phi^{(1)} \partial_{r} \partial_{s} \phi^{(1)}+\frac{2}{ \mathcal{H}} \phi^{(1)}  \partial_{r} \partial_{s} \phi^{(1)'}-\frac{1}{\mathcal{H}^{2}} \partial_{r} \phi^{(1)'} \partial_{s}  \phi^{(1)'}\right) \ .
	\end{aligned}
\end{equation}
In momentum space the equation of motion of second order scalar perturbations can be written as
\begin{equation}\label{eq:es1}
	\begin{aligned}
		\phi^{(2)}(\eta,\mathbf{p})-\psi^{(2)}(\eta,\mathbf{p})=S_{\phi}^{(2)}(\eta,\mathbf{p}) \ ,
	\end{aligned}
\end{equation}
\begin{equation}\label{eq:es2}
	\begin{aligned}
		\psi^{(2)''}(\eta,\mathbf{p})+\frac{3}{\eta} \psi^{(2)'}(\eta,\mathbf{p})+\frac{5}{6}p^2 \psi^{(2)}(\eta,\mathbf{p})+\frac{1}{\eta} \phi^{(2)'}(\eta,\mathbf{p})-\frac{1}{2} p^2 \phi^{(2)}(\eta,\mathbf{p})=\mathcal{S}^{(2)}_{\psi}(\eta,\mathbf{p}) \ .
	\end{aligned}
\end{equation}
$S_{\phi}^{(2)}(\eta,\mathbf{p}) $ and  $\mathcal{S}^{(2)}_{\psi}(\eta,\mathbf{p})$ are of the form
\begin{equation}\label{eq:Sphi}
	\begin{aligned}		
		S_{\phi}^{(2)}(\eta,\mathbf{p})&=\frac{2}{p^2}\left(\frac{3}{2}\frac{p^{r} p^{s}}{p^2}-\frac{1}{2} \delta^{rs}\right) \mathcal{S}_{rs}(\eta,\mathbf{p})\\
		&=\int \frac{\mathrm{d}^{3} q}{(2 \pi)^{3/2}}\Bigg[ \left( \frac{2(\mathbf{p}\cdot \mathbf{q})}{p^2} +\frac{9(\mathbf{p}\cdot \mathbf{q})^2}{p^4}-\frac{3q^2}{p^2} \right)\phi(\mathbf{p}- \mathbf{q}) \phi(\mathbf{q})\\
		&-\left( \frac{2}{\mathcal{H}} \right) \left(\frac{(\mathbf{p}\cdot \mathbf{q})}{p^2}-\frac{3(\mathbf{p}\cdot \mathbf{q})^2}{2p^4} +\frac{q^2}{2p^2}\right)(\phi'(\mathbf{p}- \mathbf{q})\phi(\mathbf{q})+\phi(\mathbf{p}- \mathbf{q})\phi'(\mathbf{q}) )\\
		&-\left( \frac{2}{\mathcal{H}^2} \right) \left(\frac{(\mathbf{p}\cdot \mathbf{q})}{p^2}-\frac{3(\mathbf{p}\cdot \mathbf{q})^2}{2p^4} +\frac{q^2}{2p^2}\right)\phi'(\mathbf{p}- \mathbf{q})\phi'(\mathbf{q})\Bigg] \ ,
	\end{aligned}
\end{equation}
\begin{equation}\label{eq:Spsi}
	\begin{aligned}
		\mathcal{S}^{(2)}_{\psi}(\eta,\mathbf{p})&=-\frac{1}{2} \left(\delta^{rs}-\frac{p^rp^s}{p^2} \right) \mathcal{S}^{(2)}_{rs}(\eta,\mathbf{p})\\
		&=-\int \frac{\mathrm{d}^{3} q}{(2 \pi)^{3/2}}\Bigg[\frac{11}{3} ((\mathbf{p}-\mathbf{q})\cdot\mathbf{q}) \phi^{(1)}(\mathbf{p}-\mathbf{q}) \phi^{(1)}(\mathbf{q})\\
		&+  24 \mathcal{H} \phi^{(1)} (\mathbf{p}-\mathbf{q}) \phi^{(1)'}(\mathbf{q})+\frac{16}{3}q^2 \phi^{(1)}(\mathbf{p}-\mathbf{q})  \phi^{(1)}(\mathbf{q})\\
		&-\frac{2}{3 \mathcal{H}} ((\mathbf{p}-\mathbf{q})\cdot \mathbf{q}) \phi^{(1)'}(\mathbf{p}-\mathbf{q}) \phi^{(1)}(\mathbf{q})+2 \phi^{(1)'}(\mathbf{p}-\mathbf{q})\phi^{(1)'}(\mathbf{q})\\
		&+4 \phi^{(1)}(\mathbf{p}-\mathbf{q}) \phi^{(1)''} (\mathbf{q})-\frac{1}{3\mathcal{H}^{2}} ((\mathbf{p}-\mathbf{q})\cdot\mathbf{q}) \phi^{(1)'}(\mathbf{p}-\mathbf{q})  \phi^{(1)'}(\mathbf{q})+\left( \frac{(\mathbf{p}\cdot \mathbf{q})^2}{p^2}-q^2\right)\\
		&\times\left(\frac{3}{2}\phi(\mathbf{p}- \mathbf{q})\phi(\mathbf{q})+\frac{1}{\mathcal{H}}\phi(\mathbf{p}- \mathbf{q})\phi'(\mathbf{q})+\frac{1}{2\mathcal{H}^2}\phi'(\mathbf{p}- \mathbf{q})\phi'(\mathbf{q})\right)\Bigg] \ .
	\end{aligned}
\end{equation}
Substituting $|\mathbf{p}-\mathbf{q}|=\bar{u}p$, $\mathbf{q}=\bar{v}p$, and $y=p\eta$ into Eq.~(\ref{eq:Sphi}) and Eq.~(\ref{eq:Spsi}), we obtain 
\begin{equation}
	\begin{aligned}		
		S_{\phi}^{(2)}(\bar{u},\bar{v},y)&=\int \frac{\mathrm{d}^{3} q}{(2 \pi)^{3/2}}\Phi_{\mathbf{p}-\mathbf{q}} \Phi_{\mathbf{q}}f^{(2)}_{\phi}(\bar{u},\bar{v},y)\\
		&=\int \frac{\mathrm{d}^{3} q}{(2 \pi)^{3/2}}\Phi_{\mathbf{p}-\mathbf{q}} \Phi_{\mathbf{q}}\Bigg[\left( (1+\bar{v}^2-\bar{u}^2) +\frac{9(1+\bar{v}^2-\bar{u}^2)^2}{4}-3\bar{v}^2 \right) T_{\phi}(\bar{u}y)T_{\phi}(\bar{v}y)\\
		&-y \left(1+\bar{u}^2-\bar{v}^2-\frac{3(1+\bar{v}^2-\bar{u}^2)^2}{4} +\bar{v}^2\right)(\bar{u}\frac{d}{d(\bar{u}y)}T_{\phi}(\bar{u}y)T_{\phi}(\bar{v}y)+\bar{v}T_{\phi}(\bar{u}y)\frac{d}{d(\bar{v}y)}T_{\phi}(\bar{v}y) )\\
		&-y^2 \left(1+\bar{u}^2-\bar{v}^2-\frac{3(1+\bar{v}^2-\bar{u}^2)^2}{4} +\bar{v}^2\right)\bar{u}\bar{v}\frac{d}{d(\bar{u}y)}T_{\phi}(\bar{u}y)\frac{d}{d(\bar{v}y)}T_{\phi}(\bar{v}y)\Bigg] \ ,
	\end{aligned}
\end{equation}

\begin{equation}
	\begin{aligned}
		\mathcal{S}^{(2)}_{\psi}(\bar{u},\bar{v},y)&=\int \frac{\mathrm{d}^{3} q}{(2 \pi)^{3/2}}\Phi_{\mathbf{p}-\mathbf{q}} \Phi_{\mathbf{q}}p^2f^{(2)}_{\psi}(\bar{u},\bar{v},y)\\
		&=\int \frac{\mathrm{d}^{3} q}{(2 \pi)^{3/2}}\Phi_{\mathbf{p}-\mathbf{q}} \Phi_{\mathbf{q}}p^2\Bigg[ -\frac{11(1-\bar{v}^2-\bar{u}^2)}{6} T_{\phi}(\bar{u}y) T_{\phi}(\bar{v}y)- \frac{24\bar{v}}{y} T_{\phi}(\bar{u}y) \frac{d}{d(\bar{v}y)}T_{\phi}(\bar{v}y)\\
		&-\frac{16}{3}\bar{v}^2 T_{\phi}(\bar{u}y) T_{\phi}(\bar{v}y)+\frac{\bar{u}y(1-\bar{v}^2-\bar{u}^2)}{3} \frac{d}{d(\bar{u}y)}T_{\phi}(\bar{u}y) T_{\phi}(\bar{v}y)\\
		&-2\bar{u}\bar{v} \frac{d}{d(\bar{u}y)}T_{\phi}(\bar{u}y) \frac{d}{d(\bar{v}y)}T_{\phi}(\bar{v}y)-4 \bar{v}^2T_{\phi}(\bar{u}y) \frac{d^2}{d(\bar{v}y)^2}T_{\phi}(\bar{v}y)\\
		&+\frac{\bar{u}\bar{v}y^2(1-\bar{v}^2-\bar{u}^2)}{6} \frac{d}{d(\bar{u}y)}T_{\phi}(\bar{u}y) \frac{d}{d(\bar{v}y)}T_{\phi}(\bar{v}y)-\left( \frac{(1+\bar{v}^2-\bar{u}^2)^2}{4}-\bar{v}^2 \right)\\
		&\times\left(\frac{3}{2}T_{\phi}(\bar{u}y)T_{\phi}(\bar{v}y)+\bar{v}yT_{\phi}(\bar{u}y)\frac{d}{d(\bar{v}y)}T_{\phi}(\bar{v}y)+\frac{\bar{u}\bar{v}y^2}{2}\frac{d}{d(\bar{u}y)}T_{\phi}(\bar{u}y)\frac{d}{d(\bar{v}y)}T_{\phi}(\bar{v}y)\right)\Bigg] \ .
	\end{aligned}
\end{equation}
Substituting Eq.~(\ref{eq:es1}) into Eq.~(\ref{eq:es2}), we can get the equation for $\psi^{(2)}$
\begin{equation}\label{eq:111}
	\begin{aligned}
		\psi^{(2)''}(\eta,\mathbf{p})+\frac{4}{\eta} \psi^{(2)'}(\eta,\mathbf{p})+\frac{p^2}{3} \psi^{(2)}(\eta,\mathbf{p})=\mathcal{S}^{(2)}_{r}(\eta,\mathbf{p}) \ ,
	\end{aligned}
\end{equation}
where $\mathcal{S}^{(2)}_{r}(\eta,\mathbf{p})$ is defined by
\begin{equation}
	\begin{aligned}
		\mathcal{S}^{(2)}_{r}(\eta,\mathbf{p})=\mathcal{S}^{(2)}_{\psi}(\eta,\mathbf{p})+\frac{p^2}{2}\mathcal{S}^{(2)}_{\phi}(\eta,\mathbf{p})-\frac{1}{\eta}\mathcal{S}^{(2)'}_{\phi}(\eta,\mathbf{p}) \ .
	\end{aligned}
\end{equation}
We can rewrite Eq.~(\ref{eq:111}) by introducing the new notation $z(\eta,\mathbf{p})=a^2\psi^{(2)}(\eta,\mathbf{p})$,
\begin{equation}\label{eq:ez}
	\begin{aligned}
		z^{\prime \prime}(\eta,\mathbf{p})+\left(\frac{p^{2}}{3}-\frac{2}{\eta^{2}}\right) z(\eta,\mathbf{p})=a^{2} S_{\mathrm{r}}^{(2)}(\eta,\mathbf{p}) \ .
	\end{aligned}
\end{equation}
The Green's function of Eq.~(\ref{eq:ez}) is given by
\begin{equation}
	\begin{aligned}
		p G_{\mathrm{r}}(p, \eta ; \bar{\eta})=-\Theta(\eta-\bar{\eta}) \frac{y \bar{y}}{\sqrt{3}}\left[j_{1}(y / \sqrt{3}) N_{1}(\bar{y} / \sqrt{3})-j_{1}(\bar{y} / \sqrt{3}) N_{1}(y / \sqrt{3})\right] \ ,
	\end{aligned}
\end{equation}
where $j_1$ is the spherical Bessel function of the first kind, $N_1$ is the spherical Bessel function of the second kind. The solutions of Eq.~(\ref{eq:es1}) and Eq.~(\ref{eq:es2}) are given by
\begin{equation}
	\begin{aligned}
		\Psi^{(2)}(\eta,\mathbf{p}) &=\int_{0}^{\eta} \mathrm{d} \bar{\eta}\left(\frac{a(\bar{\eta})}{a(\eta)}\right)^{2} pG_{\mathrm{r}}(p, \eta ; \bar{\eta}) S_{\mathrm{r}}^{(2)}(\eta,\mathbf{p}) =\int \frac{\mathrm{d}^{3} q}{(2 \pi)^{3/2}}  I_{\psi}^{(2)}(\bar{u},\bar{v},\bar{y})\Phi_{\mathbf{p}-\mathbf{q}} \Phi_{\mathbf{q}} \ ,
	\end{aligned}
\end{equation}
\begin{equation}
	\begin{aligned}
		\Phi^{(2)}(\eta,\mathbf{p})=\int \frac{\mathrm{d}^{3} q}{(2 \pi)^{3/2}}  I_{\phi}^{(2)}(\bar{u},\bar{v},\bar{y})\Phi_{\mathbf{p}-\mathbf{q}} \Phi_{\mathbf{q}} \ ,
	\end{aligned}
\end{equation}
where $I_{\psi}(\bar{u},\bar{v},y)$ and $I_{\phi}(\bar{u},\bar{v},y)$ are corresponding kernel functions
\begin{equation}
	\begin{aligned}
		I_{\Psi}(\bar{u},\bar{v},y) = \int_{0}^{y} \mathrm{~d} \bar{y}\left(\frac{\bar{y}}{y}\right)^{2}p G_{\mathrm{r}}(p, \eta ; \bar{\eta}) f_{\mathrm{r}}^{(2)}(\bar{u},\bar{v},\bar{y}) \ ,
	\end{aligned}
\end{equation}
\begin{equation}
	\begin{aligned}
		I_{\phi}^{(2)}(\bar{u},\bar{v},\bar{y})=I_{\psi}^{(2)}(\bar{u},\bar{v},\bar{y})+f_{\phi}(\bar{u},\bar{v},\bar{y}) \ .
	\end{aligned}
\end{equation} 

\section{Six-point correlation function}\label{sec:6}
In this appendix, we use Wick's theorem to simplify the six-point correlation function in Sec.~\ref{sec:Ph}.
We define $\mathcal{C}(\mathbf{k},\mathbf{k}',\mathbf{p},\mathbf{p}',\mathbf{q},\mathbf{q}')$ as
\begin{equation}\label{eq:C}
	\begin{aligned}
		&\mathcal{C}(\mathbf{k},\mathbf{k}',\mathbf{p},\mathbf{p}',\mathbf{q},\mathbf{q}')=\frac{1}{(2\pi^2)^3}\int{\rm d}^3\mathbf{k}' \langle \Phi_{\mathbf{k}-\mathbf{p}} \Phi_{\mathbf{p}-\mathbf{q}} \Phi_{\mathbf{q}}\Phi_{\mathbf{k}'-\mathbf{p}'} \Phi_{\mathbf{p}'-\mathbf{q}'} \Phi_{\mathbf{q}'} \rangle \\
		&=\Bigg[\left( \frac{1}{(k-p)^3(p-q)^3q^3}\delta(\mathbf{p}+\mathbf{p}')\delta(\mathbf{q}+\mathbf{q}')P_{\Phi}(k-p)P_{\Phi}(p-q)P_{\Phi}(q)\right) \\
		&+\left( \frac{1}{(k-p)^3(p-q)^3q^3}\delta(\mathbf{p}+\mathbf{p}')\delta(\mathbf{q}+\mathbf{p}'-\mathbf{q}')P_{\Phi}(k-p)P_{\Phi}(p-q)P_{\Phi}(q)\right) \\
		&+\left( \frac{1}{(k-p)^3(p-q)^3q^3}\delta(\mathbf{p}-\mathbf{q}-\mathbf{k}-\mathbf{p}')\delta(\mathbf{q}+\mathbf{q}')P_{\Phi}(k-p)P_{\Phi}(p-q)P_{\Phi}(q)\right) \\
		&+\left( \frac{1}{(k-p)^3(p-q)^3q^3}\delta(\mathbf{p}-\mathbf{q}+\mathbf{q}')\delta(\mathbf{q}-\mathbf{k}-\mathbf{p}')P_{\Phi}(k-p)P_{\Phi}(p-q)P_{\Phi}(q)\right) \\
		&+\left( \frac{1}{(k-p)^3(p-q)^3q^3}\delta(\mathbf{p}-\mathbf{k}-\mathbf{q}')\delta(\mathbf{q}+\mathbf{p}'-\mathbf{q}')P_{\Phi}(k-p)P_{\Phi}(p-q)P_{\Phi}(q)\right) \\
		&+\left( \frac{1}{(k-p)^3(p-q)^3q^3}\delta(-\mathbf{k}+\mathbf{p}-\mathbf{q}')\delta(\mathbf{q}-\mathbf{k}-\mathbf{p}')P_{\Phi}(k-p)P_{\Phi}(p-q)P_{\Phi}(q)\right) \\
		&+\left( \frac{1}{(k-p)^3q^3(p'-q')^3}\delta(\mathbf{q}-\mathbf{k})\delta(\mathbf{p}')P_{\Phi}(k-p)P_{\Phi}(q)P_{\Phi}(p'-q')\right) \\
		&+\left( \frac{1}{(k-p)^3q^3(-k-p')^3}\delta(\mathbf{q}-\mathbf{k})\delta(-\mathbf{k}-\mathbf{p}'+\mathbf{q}')P_{\Phi}(k-p)P_{\Phi}(q)P_{\Phi}(-k-p')\right) \\
		&+\left( \frac{1}{(k-p)^3q^3(-k-p')^3}\delta(\mathbf{q}+\mathbf{q}')\delta(-\mathbf{k}-\mathbf{q}')P_{\Phi}(k-p)P_{\Phi}(q)P_{\Phi}(-k-p')\right) \\
		&+ \left( \frac{1}{(k-p)^3(p-q)^3(p'-q')^3}\delta(\mathbf{p}-\mathbf{q}-\mathbf{k})\delta(\mathbf{p}')P_{\Phi}(k-p)P_{\Phi}(p-q)P_{\Phi}(p'-q')\right) \\
		&+ \left( \frac{1}{(k-p)^3(p-q)^3(-k-p')^3}\delta(\mathbf{p}-\mathbf{q}-\mathbf{k})\delta(-\mathbf{k}-\mathbf{p}'+\mathbf{q}')P_{\Phi}(k-p)P_{\Phi}(p-q)P_{\Phi}(-k-p')\right) \\
		&+\left( \frac{1}{(k-p)^3(p-q)^3(-k-p')^3}\delta(\mathbf{p}-\mathbf{q}-\mathbf{k})\delta(-\mathbf{k}-\mathbf{q}')P_{\Phi}(k-p)P_{\Phi}(p-q)P_{\Phi}(-k-p')\right) \\
		&+\left( \frac{1}{(k-p)^3(p-q)^3(p'-q')^3}\delta(\mathbf{p})\delta(\mathbf{p}')P_{\Phi}(k-p)P_{\Phi}(p-q)P_{\Phi}(p'-q')\right) \\
		&+\left( \frac{1}{(k-p)^3(p-q)^3(-k-p')^3}\delta(\mathbf{p})\delta(-\mathbf{k}-\mathbf{p}'+\mathbf{q}')P_{\Phi}(k-p)P_{\Phi}(p-q)P_{\Phi}(-k-p')\right) \\
		&+\left( \frac{1}{(k-p)^3(p-q)^3(-k-p')^3}\delta(\mathbf{p})\delta(-\mathbf{k}-\mathbf{q}')P_{\Phi}(k-p)P_{\Phi}(p-q)P_{\Phi}(-k-p')\right)\Bigg] \ .
	\end{aligned}
\end{equation}

\section{The expressions of $\mathbb{P}^{ij}(\eta, \mathbf{k})$}\label{sec:D}
Explicit expressions of $\mathbb{P}^{ij}_a$ in Eq.~(\ref{Z3}) and Eq.~(\ref{Z46}) are shown as follows,

\begin{equation}
	\begin{aligned}
&\mathbb{P}^{11}_1=\mathbb{P}^{11}	\left( \textbf{k}, \textbf{p}, - \textbf{p}, \textbf{q}, - \textbf{q}	\right) +\mathbb{P}^{11} \left( \textbf{k}, \textbf{p}, -	\textbf{p}, \textbf{q}, \textbf{q} - \textbf{p} \right) \\
&=\frac{k^4}{16 \tilde{k}^8}\left( \tilde{k}^4 \left(w^2-1\right)^2-2 \tilde{k}^2 \left(w^2+1\right)+1\right) \left(\tilde{k}^4 \left(v^2+w^2\right)^2-4 \tilde{k}^2 \left(v^2+w^2+1\right)+4\right) \ ,
\end{aligned}
\end{equation}

\begin{equation}
	\begin{aligned}
&\mathbb{P}^{11}_2=\mathbb{P}^{11} \left( \textbf{k}, \textbf{p}, \textbf{q} -	\textbf{k}, \textbf{q}, \textbf{q} - \textbf{p} \right) +\mathbb{P}^{11} \left( \textbf{k}, \textbf{p}, \textbf{q} - \textbf{k},	\textbf{q}, - \textbf{k} + \textbf{p} \right)\\
&=\frac{k^4}{16 \tilde{k}^8}\left(\tilde{k}^4 \left(v^2+w^2\right)^2-4 \tilde{k}^2 \left(v^2+w^2+1\right)+4\right) \left(\tilde{k}^4 \left(v^2 \left(w^2+1\right)+w^2-1\right)-\tilde{k}^2 \left(v^2+w^2\right)+1\right) \ ,
\end{aligned}
\end{equation}

\begin{equation}
	\begin{aligned}
		&\mathbb{P}^{11}_3=\mathbb{P}^{11} \left( \textbf{k}, \textbf{p}, \textbf{p}-\textbf{q} -	\textbf{k}, \textbf{q}, -\textbf{q} \right) +\mathbb{P}^{11} \left( \textbf{k}, \textbf{p}, \textbf{p}-\textbf{q} - \textbf{k},	\textbf{q}, \textbf{p}- \textbf{k}   \right)\\
		&=\frac{k^4}{16 \tilde{k}^8}\left(\tilde{k}^4 \left(w^2-1\right)^2-2 \tilde{k}^2 \left(w^2+1\right)+1\right) \left(\tilde{k}^4 \left(v^4+v^2 \left(w^2-1\right)+w^2\right)-\tilde{k}^2 \left(3 v^2+w^2+2\right)+2\right) \ ,
	\end{aligned}
\end{equation}

\begin{equation}
	\begin{aligned}
		&\mathbb{P}^{22}_1=\mathbb{P}^{22}	\left( \textbf{k}, \textbf{p}, - \textbf{p}, \textbf{q}, - \textbf{q}	\right) +\mathbb{P}^{22} \left( \textbf{k}, \textbf{p}, -	\textbf{p}, \textbf{q}, \textbf{q} - \textbf{p} \right) \\
		&=\frac{k^4}{1024 \tilde{k}^{12} v^4}\left(16 \tilde{k}^8 v^4 \left(\tilde{k}^2 v^2-4\right)^2+8 \tilde{k}^4 v^2 \left(\tilde{k}^2 v^2-4\right) \left(\tilde{k}^6 v^2 \left(3 v^4+v^2 \left(8 w^2-2\right)+8 w^4-8 w^2+3\right) \right.\right.\\
		&\left.\left.-2 \tilde{k}^4 \left(9 v^4+v^2 \left(12 w^2-1\right)+2\right)+\tilde{k}^2 \left(27 v^2+8\right)-4\right)+\left(\tilde{k}^6 v^2 \left(v^4+v^2 \left(8 w^2-6\right)+8 w^4-8 w^2+1\right) \right.\right.\\
		&\left.\left.+\tilde{k}^4 \left(-6 v^4+v^2 \left(22-24 w^2\right)+4\right)+\tilde{k}^2 \left(9 v^2-8\right)+4\right)^2 \right)\ ,
	\end{aligned}
\end{equation}

\begin{equation}
	\begin{aligned}
		&\mathbb{P}^{22}_2=\mathbb{P}^{22} \left( \textbf{k}, \textbf{p}, \textbf{q} -	\textbf{k}, \textbf{q}, \textbf{q} - \textbf{p} \right) +\mathbb{P}^{22} \left( \textbf{k}, \textbf{p}, \textbf{q} - \textbf{k},	\textbf{q}, - \textbf{k} + \textbf{p} \right)\\
		&=\frac{k^4}{1024 \tilde{k}^{12} v^2 w^2} \left( \tilde{k}^{12} v^2 w^2 \left(8 v^8+8 v^6 \left(9 w^2+5\right)+v^4 \left(129 w^4+190 w^2-39\right)+2 v^2 \left(36 w^6\right.\right.\right. \\
		&\left.\left.\left.+95 w^4-74 w^2-1\right)+8 w^8+40 w^6-39 w^4-2 w^2+1\right)+\tilde{k}^{10} \left(-72 v^8 w^2+v^6 \left(-438 w^4-298 w^2+4\right)\right.\right. \\
		&\left.\left.+v^4 \left(-438 w^6-832 w^4+54 w^2+24\right)+v^2 \left(-72 w^8-298 w^6+54 w^4+148 w^2+4\right)+4 \left(w^6+6 w^4+w^2\right)\right)\right.\\
		&\left.+\tilde{k}^8 \left(v^6 \left(225 w^2-8\right)+18 v^4 \left(42 w^4+37 w^2-4\right)+v^2 \left(225 w^6+666 w^4-26 w^2-80\right) \right.\right. \\
		&\left.\left.-8 \left(w^6+9 w^4+10 w^2-2\right)\right)+2 \tilde{k}^6 \left(2 v^6+v^4 \left(36-103 w^2\right)+v^2 \left(-103 w^4-134 w^2+92\right)\right.\right.\\
		&\left.\left.+2 \left(w^6+18 w^4+46 w^2+16\right)\right)-\tilde{k}^4 \left(24 v^4+3 v^2 \left(37 w^2+48\right)+8 \left(3 w^4+18 w^2+20\right)\right) \right.\\
		&\left.+4 \tilde{k}^2 \left(9 v^2+9 w^2+16\right)+16 \right) \ ,
	\end{aligned}
\end{equation}

\begin{equation}
	\begin{aligned}
		&\mathbb{P}^{22}_3=\mathbb{P}^{22} \left( \textbf{k}, \textbf{p}, \textbf{p}-\textbf{q} -	\textbf{k}, \textbf{q}, -\textbf{q} \right) +\mathbb{P}^{22} \left( \textbf{k}, \textbf{p}, \textbf{p}-\textbf{q} - \textbf{k},	\textbf{q}, \textbf{p}- \textbf{k}   \right)\\
		&=\frac{k^4}{1024 \tilde{k}^{10} v^2 \left(\tilde{k}^2 \left(v^2+w^2-1\right)-3\right)} \left( \tilde{k}^{12} v^2 \left(v^{10}+3 v^8 \left(w^2-1\right)+v^6 \left(-37 w^4+144 w^2-37\right) \right.\right. \\
		&\left.\left.+v^4 \left(-79 w^6+377 w^4-377 w^2+79\right)+v^2 \left(-32 w^8+158 w^6-251 w^4+158 w^2-32\right)+8 w^{10} \right.\right. \\
		&\left.\left.-80 w^8+201 w^6-201 w^4+80 w^2-8\right)+\tilde{k}^{10} \left(-9 v^{10}+26 v^8+v^6 \left(267 w^4-902 w^2+215\right) \right.\right. \\
		&\left.\left.+2 v^4 \left(117 w^6-517 w^4+569 w^2-122\right)+v^2 \left(-48 w^8+374 w^6-525 w^4+208 w^2-12\right)\right.\right. \\
		&\left.\left.+4 \left(w^6-9 w^4+16 w^2-8\right)\right)+\tilde{k}^8 \left(27 v^8-3 v^6 \left(27 w^2+25\right)+v^4 \left(-459 w^4+1410 w^2-428\right) \right.\right.\\
		&\left.\left.+v^2 \left(81 w^6-411 w^4-104 w^2+152\right)-4 \left(2 w^6-15 w^4+8 w^2+12\right)\right)+\tilde{k}^6 \left(-27 v^6+10 v^4 \left(11 w^2 \right.\right.\right. \\
		&\left.\left.\left.+18\right)+v^2 \left(-79 w^4+160 w^2+240\right)+4 \left(w^6-3 w^4-32 w^2+24\right)\right)+4 \tilde{k}^4 \left(3 v^4+42 v^2 \left(w^2-2\right)\right.\right. \\
		&\left.\left.-3 w^4+24 w^2+16\right)-4 \tilde{k}^2 \left(9 v^2+16\right)-16 \right)\ ,
	\end{aligned}
\end{equation}

\begin{equation}
	\begin{aligned}
		\mathbb{P}^{33}_1=\mathbb{P}^{33}	\left( \textbf{k}, \textbf{p}, - \textbf{p}, \textbf{q}, - \textbf{q}	\right) +\mathbb{P}^{33} \left( \textbf{k}, \textbf{p}, -	\textbf{p}, \textbf{q}, \textbf{q} - \textbf{p} \right) = 0\ ,
	\end{aligned}
\end{equation}

\begin{equation}
	\begin{aligned}
		\mathbb{P}^{33}_2=\mathbb{P}^{33} \left( \textbf{k}, \textbf{p}, \textbf{q} -	\textbf{k}, \textbf{q}, \textbf{q} - \textbf{p} \right) +\mathbb{P}^{33} \left( \textbf{k}, \textbf{p}, \textbf{q} - \textbf{k},	\textbf{q}, - \textbf{k} + \textbf{p} \right)= 0\ ,
	\end{aligned}
\end{equation}

\begin{equation}
	\begin{aligned}
		\mathbb{P}^{33}_3=\mathbb{P}^{33} \left( \textbf{k}, \textbf{p}, \textbf{p}-\textbf{q} -	\textbf{k}, \textbf{q}, -\textbf{q} \right) +\mathbb{P}^{33} \left( \textbf{k}, \textbf{p}, \textbf{p}-\textbf{q} - \textbf{k},	\textbf{q}, \textbf{p}- \textbf{k}   \right)=0 \ ,
	\end{aligned}
\end{equation}

\begin{equation}
	\begin{aligned}
		&\mathbb{P}^{44}_1=\mathbb{P}^{44}	\left( \textbf{k}, \textbf{p}, - \textbf{p}, \textbf{q}, - \textbf{q}	\right) +\mathbb{P}^{44} \left( \textbf{k}, \textbf{p}, -	\textbf{p}, \textbf{q}, \textbf{q} - \textbf{p} \right)=\frac{k^4}{16}\left(\left(-\frac{1}{\tilde{k}^2}+v^2+1\right)^2-4 v^2\right)^2 \ ,
	\end{aligned}
\end{equation}

\begin{equation}
	\begin{aligned}
		&\mathbb{P}^{44}_2=\mathbb{P}^{44} \left( \textbf{k}, \textbf{p}, \textbf{q} -	\textbf{k}, \textbf{q}, \textbf{q} - \textbf{p} \right) +\mathbb{P}^{44} \left( \textbf{k}, \textbf{p}, \textbf{q} - \textbf{k},	\textbf{q}, - \textbf{k} + \textbf{p} \right)\\
		&=\frac{k^4}{16 }\left( 4 v^2 \left(\frac{1}{\text{ks}^2}-w^2+1\right)^2-8 v^2 \left(-\frac{1}{\tilde{k}^2}+v^2+1\right) \left(\frac{1}{\tilde{k}^2}-w^2+1\right)\right. \\
		&\left.+\left(-\frac{1}{\tilde{k}^2}+v^2+1\right)^2   \left(\frac{1}{\tilde{k}^2}-w^2+1\right)^2-\frac{16 v^2}{\tilde{k}^2}+\frac{4 \left(-\frac{1}{\tilde{k}^2}+v^2+1\right)^2}{\tilde{k}^2}+8 v^4 \right) \ ,
	\end{aligned}
\end{equation}

\begin{equation}
	\begin{aligned}
		&\mathbb{P}^{44}_3=\mathbb{P}^{44} \left( \textbf{k}, \textbf{p}, \textbf{p}-\textbf{q} -	\textbf{k}, \textbf{q}, -\textbf{q} \right) +\mathbb{P}^{44} \left( \textbf{k}, \textbf{p}, \textbf{p}-\textbf{q} - \textbf{k},	\textbf{q}, \textbf{p}- \textbf{k}   \right)\\
		&=\frac{k^4}{16 \tilde{k}^8}\left( \tilde{k}^8 \left(v^8+2 v^6 \left(w^2-1\right)+v^4 \left(w^4+4 w^2+1\right)+6 v^2 w^2 \left(w^2-1\right)+w^4\right)\right. \\
		&\left.-2 \tilde{k}^6 \left(3 v^6+v^4 \left(4 w^2-1\right)+v^2 \left(w^4+10 w^2-2\right)+w^4+2 w^2-2\right)+\tilde{k}^4 \left(13 v^4+2 v^2 \left(5 w^2+4\right)\right.\right.\\
		&\left.\left.+w^4+8 w^2-4\right)-4 \tilde{k}^2 \left(3 v^2+w^2+1\right)+4 \right) \ ,
	\end{aligned}
\end{equation}

\begin{equation}
	\begin{aligned}
		&\mathbb{P}^{12}_1=\mathbb{P}^{12}	\left( \textbf{k}, \textbf{p}, - \textbf{p}, \textbf{q}, - \textbf{q}	\right) +\mathbb{P}^{12} \left( \textbf{k}, \textbf{p}, -	\textbf{p}, \textbf{q}, \textbf{q} - \textbf{p} \right)\\
		&=-\frac{k^4}{128 \tilde{k}^{10} v^2}\left( \tilde{k}^{10} \left(v^8 \left(w^2+1\right)+v^6 \left(9 w^4-11 w^2+6\right)+v^4 \left(16 w^6-26 w^4+11 w^2+1\right) \right.\right. \\
		&\left.\left.+v^2 w^2 \left(8 w^6-16 w^4+9 w^2-1\right)\right)-\tilde{k}^8 \left(v^8+v^6 \left(17 w^2+14\right)+v^4 \left(62 w^4-34 w^2+39\right) \right.\right.\\
		&\left.\left.+v^2 \left(48 w^6-38 w^4-11 w^2+22\right)-4 w^2 \left(w^2-1\right)\right)+\tilde{k}^6 \left(8 v^6+v^4 \left(67 w^2+59\right)+v^2 \left(97 w^4-3 w^2+90\right)\right.\right. \\
		&\left.\left.-4 \left(2 w^4+w^2-6\right)\right)-\tilde{k}^4 \left(21 v^4+v^2 \left(71 w^2+82\right)-4 \left(w^4+5 w^2-10\right)\right)\right.\\
		&\left.+2 \tilde{k}^2 \left(7 v^2-6 w^2+4\right)+8 \right) \ ,
	\end{aligned}
\end{equation}

\begin{equation}
	\begin{aligned}
		&\mathbb{P}^{12}_2=\mathbb{P}^{12} \left( \textbf{k}, \textbf{p}, \textbf{q} -	\textbf{k}, \textbf{q}, \textbf{q} - \textbf{p} \right) +\mathbb{P}^{12} \left( \textbf{k}, \textbf{p}, \textbf{q} - \textbf{k},	\textbf{q}, - \textbf{k} + \textbf{p} \right)\\
		&=-\frac{k^4}{128 \tilde{k}^{10} w^2}\left( \tilde{k}^{10} w^2 \left(8 v^6 \left(w^2+1\right)+8 v^4 \left(2 w^4+3 w^2-1\right)+v^2 \left(9 w^6+23 w^4-17 w^2+1\right)\right.\right. \\
		&\left.\left.+w^2 \left(w^6+5 w^4-5 w^2-1\right)\right)-\tilde{k}^8 \left(8 v^6 w^2+8 v^4 w^2 \left(7 w^2+4\right)+v^2 \left(55 w^6+102 w^4+7 w^2-4\right)\right.\right.\\
		&\left.\left.+w^2 \left(9 w^6+40 w^4-15 w^2-26\right)\right)+\tilde{k}^6 \left(40 v^4 w^2+v^2 \left(103 w^4+59 w^2-12\right)+29 w^6+85 w^4-14 w^2-8\right)\right. \\
		&\left.+\tilde{k}^4 \left(v^2 \left(12-53 w^2\right)-35 w^4-18 w^2+24\right)+\tilde{k}^2 \left(6 \left(w^2-4\right)-4 v^2\right)+8 \right) \ ,
	\end{aligned}
\end{equation}

\begin{equation}
	\begin{aligned}
		&\mathbb{P}^{12}_3=\mathbb{P}^{12} \left( \textbf{k}, \textbf{p}, \textbf{p}-\textbf{q} -	\textbf{k}, \textbf{q}, -\textbf{q} \right) +\mathbb{P}^{12} \left( \textbf{k}, \textbf{p}, \textbf{p}-\textbf{q} - \textbf{k},	\textbf{q}, \textbf{p}- \textbf{k}   \right)\\
		&=-\frac{k^4}{128 \tilde{k}^8 \left(\tilde{k}^2 \left(v^2+w^2-1\right)-3\right)}\left(\tilde{k}^{10} \left(v^8 \left(w^2+1\right)+v^6 \left(-4 w^4+17 w^2-9\right)+v^4 \left(-10 w^6+45 w^4\right.\right. \right.\\
		&\left.\left.\left.-49 w^2+16\right)+v^2 \left(-4 w^8+19 w^6-31 w^4+24 w^2-8\right)+w^2 \left(w^2-1\right)^2 \left(w^4-8 w^2+8\right)\right) \right. \\
		&\left.-\tilde{k}^8 \left(v^8+v^6 \left(w^2-4\right)+v^4 \left(-35 w^4+108 w^2-18\right)+v^2 \left(-29 w^6+132 w^4-128 w^2+12\right)+6 w^8 \right.\right.\\
		&\left.\left.-44 w^6+66 w^4-44 w^2+16\right)+\tilde{k}^6 \left(5 v^6-v^4 \left(19 w^2+28\right)+v^2 \left(-61 w^4+200 w^2-12\right)\right.\right. \\
		&\left.\left.+11 w^6-28 w^4-36 w^2-8\right)+\tilde{k}^4 \left(-6 v^4+4 v^2 \left(8 w^2+7\right)-10 w^4-28 w^2+56\right)\right. \\
		&\left.+4 \tilde{k}^2 \left(v^2+3 w^2-6\right)-8\right) \ ,
	\end{aligned}
\end{equation}

\begin{equation}
	\begin{aligned}
		&\mathbb{P}^{21}_1=\mathbb{P}^{21}	\left( \textbf{k}, \textbf{p}, - \textbf{p}, \textbf{q}, - \textbf{q}	\right) +\mathbb{P}^{21} \left( \textbf{k}, \textbf{p}, -	\textbf{p}, \textbf{q}, \textbf{q} - \textbf{p} \right) \\
		&=-\frac{k^4}{128 \tilde{k}^{10} v^2}\left( \tilde{k}^{10} \left(v^8 \left(w^2+1\right)+v^6 \left(9 w^4-11 w^2+6\right)+v^4 \left(16 w^6-26 w^4+11 w^2+1\right)+v^2 w^2 \left(8 w^6 \right.\right.\right. \\
		&\left.\left.\left.-16 w^4+9 w^2-1\right)\right)-\tilde{k}^8 \left(v^8+v^6 \left(17 w^2+14\right)+v^4 \left(62 w^4-34 w^2+39\right)+v^2 \left(48 w^6-38 w^4-11 w^2\right.\right.\right.  \\
		&\left.\left.\left.+22\right)-4 w^2 \left(w^2-1\right)\right)+\tilde{k}^6 \left(8 v^6+v^4 \left(67  w^2+59\right)+v^2 \left(97 w^4-3 w^2+90\right)-4 \left(2 w^4+w^2-6\right)\right) \right. \\
		&\left.-\tilde{k}^4 \left(21 v^4+v^2 \left(71 w^2+82\right)-4 \left(w^4+5 w^2-10\right)\right)+2 \tilde{k}^2 \left(7 v^2-6 w^2+4\right)+8 \right) \ ,
	\end{aligned}
\end{equation}

\begin{equation}
	\begin{aligned}
		&\mathbb{P}^{21}_2=\mathbb{P}^{21} \left( \textbf{k}, \textbf{p}, \textbf{q} -	\textbf{k}, \textbf{q}, \textbf{q} - \textbf{p} \right) +\mathbb{P}^{21} \left( \textbf{k}, \textbf{p}, \textbf{q} - \textbf{k},	\textbf{q}, - \textbf{k} + \textbf{p} \right)\\
		&=-\frac{k^4}{128 \tilde{k}^{10} v^2}\left( \tilde{k}^{10} v^2 \left(v^8+v^6 \left(9 w^2+5\right)+v^4 \left(16 w^4+23 w^2-5\right)+v^2 \left(8 w^6+24 w^4-17 w^2-1\right)+8 w^6 \right.\right. \\
		&\left.\left.-8 w^4+w^2\right)-\tilde{k}^8 \left(9 v^8+5 v^6 \left(11 w^2+8\right)+v^4 \left(56 w^4+102 w^2-15\right)+v^2 \left(8 w^6+32 w^4+7 w^2-26\right)-4 w^2\right) \right. \\
		&\left.+\tilde{k}^6 \left(29 v^6+v^4 \left(103 w^2+85\right)+v^2 \left(40 w^4+59 w^2-14\right)-4 \left(3 w^2+2\right)\right)+\tilde{k}^4 \left(-35 v^4-v^2 \left(53 w^2+18\right) \right.\right.  \\
		&\left.\left.+12 \left(w^2+2\right)\right)+\tilde{k}^2 \left(6 v^2-4 \left(w^2+6\right)\right)+8 \right) \ ,
	\end{aligned}
\end{equation}

\begin{equation}
	\begin{aligned}
		&\mathbb{P}^{21}_3=\mathbb{P}^{21} \left( \textbf{k}, \textbf{p}, \textbf{p}-\textbf{q} -	\textbf{k}, \textbf{q}, -\textbf{q} \right) +\mathbb{P}^{21} \left( \textbf{k}, \textbf{p}, \textbf{p}-\textbf{q} - \textbf{k},	\textbf{q}, \textbf{p}- \textbf{k}   \right)\\
		&=\frac{k^4}{128 \tilde{k}^{10} v^2} \left( \tilde{k}^{10} v^2 \left(v^6 \left(w^2+1\right)+v^4 \left(8 w^4-17 w^2+5\right)+v^2 \left(8 w^6-24 w^4+23 w^2-5\right)+8 w^6-16 w^4 \right.\right.  \\
		&\left.\left.+9 w^2-1\right)-\tilde{k}^8 \left(v^8+3 v^6 \left(5 w^2+4\right)+v^4 \left(40 w^4-42 w^2+21\right)+v^2 \left(8 w^6+16 w^4-w^2+2\right)-4 w^2+4\right) \right. \\
		&\left.+\tilde{k}^6 \left(7 v^6+v^4 \left(47 w^2+41\right)+v^2 \left(32 w^4+19 w^2+44\right)-12 w^2+8\right)-\tilde{k}^4 \left(15 v^4+v^2 \left(29 w^2+46\right)-12 w^2\right)\right.  \\
		&\left.+\tilde{k}^2 \left(5 v^2-4 \left(w^2+2\right)\right)+4 \right)\ ,
	\end{aligned}
\end{equation}

\begin{equation}
	\begin{aligned}
		\mathbb{P}^{13}_1=\mathbb{P}^{13}	\left( \textbf{k}, \textbf{p}, - \textbf{p}, \textbf{q}, - \textbf{q}	\right) +\mathbb{P}^{13} \left( \textbf{k}, \textbf{p}, -	\textbf{p}, \textbf{q}, \textbf{q} - \textbf{p} \right) = 0\ ,
	\end{aligned}
\end{equation}

\begin{equation}
	\begin{aligned}
		\mathbb{P}^{13}_2=\mathbb{P}^{13} \left( \textbf{k}, \textbf{p}, \textbf{q} -	\textbf{k}, \textbf{q}, \textbf{q} - \textbf{p} \right) +\mathbb{P}^{13} \left( \textbf{k}, \textbf{p}, \textbf{q} - \textbf{k},	\textbf{q}, - \textbf{k} + \textbf{p} \right)= 0\ ,
	\end{aligned}
\end{equation}

\begin{equation}
	\begin{aligned}
		\mathbb{P}^{13}_3=\mathbb{P}^{13} \left( \textbf{k}, \textbf{p}, \textbf{p}-\textbf{q} -	\textbf{k}, \textbf{q}, -\textbf{q} \right) +\mathbb{P}^{13} \left( \textbf{k}, \textbf{p}, \textbf{p}-\textbf{q} - \textbf{k},	\textbf{q}, \textbf{p}- \textbf{k}   \right)=0 \ ,
	\end{aligned}
\end{equation}

\begin{equation}
	\begin{aligned}
		&\mathbb{P}^{31}_1=\mathbb{P}^{31}	\left( \textbf{k}, \textbf{p}, - \textbf{p}, \textbf{q}, - \textbf{q}	\right) +\mathbb{P}^{31} \left( \textbf{k}, \textbf{p}, -	\textbf{p}, \textbf{q}, \textbf{q} - \textbf{p} \right) \\
		&= \frac{k^4}{32\tilde{k}^8}\left(\tilde{k}^2 \left(v^2+1\right)-1\right) \left(\tilde{k}^2 \left(v^2+2 w^2-1\right)-3\right) \left(\tilde{k}^4 \left(v^2 \left(w^2+1\right)+w^2 \left(w^2-1\right)\right)\right. \\
		&\left.-\tilde{k}^2 \left(v^2+3 w^2+2\right)+2\right)\ ,
	\end{aligned}
\end{equation}

\begin{equation}
	\begin{aligned}
		&\mathbb{P}^{31}_2=\mathbb{P}^{31} \left( \textbf{k}, \textbf{p}, \textbf{q} -	\textbf{k}, \textbf{q}, \textbf{q} - \textbf{p} \right) +\mathbb{P}^{31} \left( \textbf{k}, \textbf{p}, \textbf{q} - \textbf{k},	\textbf{q}, - \textbf{k} + \textbf{p} \right) \\
		&= \frac{k^4}{32\tilde{k}^8}\left(\tilde{k}^4 \left(v^2-1\right)^2-2 \tilde{k}^2 \left(v^2+1\right)+1\right) \left(\tilde{k}^4 \left(v^4+3 v^2 w^2+v^2+2 w^4-w^2\right)  \right. \\
		&\left. -\tilde{k}^2 \left(5 v^2+7 w^2+6\right)+6\right)\ ,
	\end{aligned}
\end{equation}

\begin{equation}
	\begin{aligned}
		&\mathbb{P}^{31}_3=\mathbb{P}^{31} \left( \textbf{k}, \textbf{p}, \textbf{p}-\textbf{q} -	\textbf{k}, \textbf{q}, -\textbf{q} \right) +\mathbb{P}^{31} \left( \textbf{k}, \textbf{p}, \textbf{p}-\textbf{q} - \textbf{k},	\textbf{q}, \textbf{p}- \textbf{k}   \right) \\
		&=-\frac{k^4}{32\tilde{k}^8}\left(\tilde{k}^4 \left(v^2-1\right)^2-2 \tilde{k}^2 \left(v^2+1\right)+1\right) \left(\tilde{k}^4 \left(v^2 \left(w^2+1\right)+2 w^4-3 w^2+1\right)\right. \\
		&\left.-\tilde{k}^2 \left(v^2+5 w^2+4\right)+3\right)\ ,
	\end{aligned}
\end{equation}

\begin{equation}
	\begin{aligned}
		&\mathbb{P}^{14}_1=\mathbb{P}^{14}	\left( \textbf{k}, \textbf{p}, - \textbf{p}, \textbf{q}, - \textbf{q}	\right) +\mathbb{P}^{14} \left( \textbf{k}, \textbf{p}, -	\textbf{p}, \textbf{q}, \textbf{q} - \textbf{p} \right) \\
		&=- \frac{k^4}{16\tilde{k}^8} \left( \tilde{k}^8 \left(v^6 \left(w^2+1\right)+v^4 \left(w^4+5 w^2-2\right)+v^2 \left(6 w^4-5 w^2+1\right)+w^2 \left(w^2-1\right)\right) \right.  \\
		&\left.-\tilde{k}^6 \left(v^6+v^4 \left(5 w^2+2\right)+v^2 \left(2 w^4+18 w^2-1\right)+2 w^4+w^2-6\right)+\tilde{k}^4 \left(4 v^4+v^2 \left(7 w^2+3\right)+w^4  \right.\right. \\
		&\left.\left.+5 w^2-10\right)+\tilde{k}^2 \left(-5 v^2-3 w^2+2\right)+2 \right) \ ,
	\end{aligned}
\end{equation}

\begin{equation}
	\begin{aligned}
		&\mathbb{P}^{14}_2=\mathbb{P}^{14} \left( \textbf{k}, \textbf{p}, \textbf{q} -	\textbf{k}, \textbf{q}, \textbf{q} - \textbf{p} \right) +\mathbb{P}^{14} \left( \textbf{k}, \textbf{p}, \textbf{q} - \textbf{k},	\textbf{q}, - \textbf{k} + \textbf{p} \right) \\
		&= -\frac{k^4}{16\tilde{k}^8}\left(\tilde{k}^4 \left(w^2-1\right)^2-2 \tilde{k}^2 \left(w^2+1\right)+1\right) \left(\tilde{k}^4 \left(v^2 \left(w^2+1\right)+w^2 \left(w^2-1\right)\right) \right.\\
		&\left.-\tilde{k}^2 \left(v^2+3 w^2+2\right)+2\right)\ ,
	\end{aligned}
\end{equation}

\begin{equation}
	\begin{aligned}
		&\mathbb{P}^{14}_3=\mathbb{P}^{14} \left( \textbf{k}, \textbf{p}, \textbf{p}-\textbf{q} -	\textbf{k}, \textbf{q}, -\textbf{q} \right) +\mathbb{P}^{14} \left( \textbf{k}, \textbf{p}, \textbf{p}-\textbf{q} - \textbf{k},	\textbf{q}, \textbf{p}- \textbf{k}   \right) \\
		&=-\frac{k^4}{16\tilde{k}^8} \left(\tilde{k}^4 \left(v^2+w^2\right)^2-4 \tilde{k}^2 \left(v^2+w^2+1\right)+4\right) \left(\tilde{k}^4 \left(v^2 \left(w^2+1\right)+w^2 \left(w^2-1\right)\right)\right. \\
		&\left.-\tilde{k}^2 \left(v^2+3 w^2+2\right)+2\right)  \ ,
	\end{aligned}
\end{equation}

\begin{equation}
	\begin{aligned}
		&\mathbb{P}^{41}_1=\mathbb{P}^{41}	\left( \textbf{k}, \textbf{p}, - \textbf{p}, \textbf{q}, - \textbf{q}	\right) +\mathbb{P}^{41} \left( \textbf{k}, \textbf{p}, -	\textbf{p}, \textbf{q}, \textbf{q} - \textbf{p} \right) \\
		&=- \frac{k^4}{16\tilde{k}^8} \left( \tilde{k}^8 \left(v^6 \left(w^2+1\right)+v^4 \left(w^4+5 w^2-2\right)+v^2 \left(6 w^4-5 w^2+1\right)+w^2 \left(w^2-1\right)\right)-\tilde{k}^6 \left(v^6+v^4 \left(5 w^2 \right.\right.\right.  \\
		&\left.\left.\left.+2\right)+v^2 \left(2 w^4+18 w^2-1\right)+2 w^4+w^2-6\right)+\tilde{k}^4 \left(4 v^4+v^2 \left(7 w^2+3\right)+w^4+5 w^2-10\right)\right.  \\
		&\left.+\tilde{k}^2 \left(-5 v^2-3 w^2+2\right)+2 \right) \ ,
	\end{aligned}
\end{equation}

\begin{equation}
	\begin{aligned}
		&\mathbb{P}^{41}_2=\mathbb{P}^{41} \left( \textbf{k}, \textbf{p}, \textbf{q} -	\textbf{k}, \textbf{q}, \textbf{q} - \textbf{p} \right) +\mathbb{P}^{41} \left( \textbf{k}, \textbf{p}, \textbf{q} - \textbf{k},	\textbf{q}, - \textbf{k} + \textbf{p} \right) \\
		&= -\frac{k^4}{16\tilde{k}^8}\left(\tilde{k}^4 \left(v^2-1\right)^2-2 \tilde{k}^2 \left(v^2+1\right)+1\right) \left(\tilde{k}^4 \left(v^4+v^2 \left(w^2-1\right)+w^2\right)-\tilde{k}^2 \left(3 v^2+w^2+2\right)+2\right)\ ,
	\end{aligned}
\end{equation}

\begin{equation}
	\begin{aligned}
		&\mathbb{P}^{41}_3=\mathbb{P}^{41} \left( \textbf{k}, \textbf{p}, \textbf{p}-\textbf{q} -	\textbf{k}, \textbf{q}, -\textbf{q} \right) +\mathbb{P}^{41} \left( \textbf{k}, \textbf{p}, \textbf{p}-\textbf{q} - \textbf{k},	\textbf{q}, \textbf{p}- \textbf{k}   \right) \\
		&=\frac{k^4}{16\tilde{k}^8} \left(\tilde{k}^4 \left(v^2-1\right)^2-2 \tilde{k}^2 \left(v^2+1\right)+1\right) \left(\tilde{k}^4 \left(v^2 \left(w^2+1\right)+w^2-1\right)-\tilde{k}^2 \left(v^2+w^2\right)+1\right)  \ ,
	\end{aligned}
\end{equation}

\begin{equation}
	\begin{aligned}
		&\mathbb{P}^{23}_1=-\mathbb{P}^{23}	\left( \textbf{k}, \textbf{p}, - \textbf{p}, \textbf{q}, - \textbf{q}	\right) +\mathbb{P}^{23} \left( \textbf{k}, \textbf{p}, -	\textbf{p}, \textbf{q}, \textbf{q} - \textbf{p} \right) =0 \ ,
	\end{aligned}
\end{equation}

\begin{equation}
	\begin{aligned}
		&\mathbb{P}^{23}_2=\mathbb{P}^{23} \left( \textbf{k}, \textbf{p}, \textbf{q} -	\textbf{k}, \textbf{q}, \textbf{q} - \textbf{p} \right) +\mathbb{P}^{23} \left( \textbf{k}, \textbf{p}, \textbf{q} - \textbf{k},	\textbf{q}, - \textbf{k} + \textbf{p} \right) = 0\ ,
	\end{aligned}
\end{equation}

\begin{equation}
	\begin{aligned}
		&\mathbb{P}^{23}_3=\mathbb{P}^{23} \left( \textbf{k}, \textbf{p}, \textbf{p}-\textbf{q} -	\textbf{k}, \textbf{q}, -\textbf{q} \right) +\mathbb{P}^{23} \left( \textbf{k}, \textbf{p}, \textbf{p}-\textbf{q} - \textbf{k},	\textbf{q}, \textbf{p}- \textbf{k}   \right) =0 \ ,
	\end{aligned}
\end{equation}

\begin{equation}
	\begin{aligned}
		&\mathbb{P}^{32}_1=\mathbb{P}^{32}	\left( \textbf{k}, \textbf{p}, - \textbf{p}, \textbf{q}, - \textbf{q}	\right) +\mathbb{P}^{32} \left( \textbf{k}, \textbf{p}, -	\textbf{p}, \textbf{q}, \textbf{q} - \textbf{p} \right) \\
		&=-\frac{k^4}{256 \tilde{k}^{10} v^2}\left(\tilde{k}^2 \left(v^2+1\right)-1\right) \left(\tilde{k}^2 \left(v^2+2 w^2-1\right)-3\right) \left(\tilde{k}^6 v^2 \left(v^4+v^2 \left(8 w^2+6\right)+8 w^4-8 w^2+1\right)\right.  \\
		&\left.-2 \tilde{k}^4 \left(3 v^4+v^2 \left(12 w^2+13\right)-2\right)+\tilde{k}^2 \left(9 v^2-8\right)+4\right) \ ,
	\end{aligned}
\end{equation}

\begin{equation}
	\begin{aligned}
		&\mathbb{P}^{32}_2=\mathbb{P}^{32} \left( \textbf{k}, \textbf{p}, \textbf{q} -	\textbf{k}, \textbf{q}, \textbf{q} - \textbf{p} \right) +\mathbb{P}^{32} \left( \textbf{k}, \textbf{p}, \textbf{q} - \textbf{k},	\textbf{q}, - \textbf{k} + \textbf{p} \right) \\
		&=-\frac{k^4}{256 \tilde{k}^{10} w^2}\left( \tilde{k}^{10} w^2 \left(8 v^8+8 v^6 \left(3 w^2-1\right)+v^4 \left(17 w^4-50 w^2-7\right)+2 v^2 \left(w^6-14 w^4+17 w^2+4\right)  \right.\right. \\
		&\left.\left.+2 w^6+11 w^4-4 w^2-1\right)-2 \tilde{k}^8 \left(28 v^6 w^2+v^4 \left(51 w^4-11 w^2-2\right)+2 v^2 \left(8 w^4-19 w^2-3\right) w^2+w^8 \right.\right. \\
		&\left.\left.+13 w^6+30 w^4+12 w^2+2\right)+\tilde{k}^6 \left(v^4 \left(129 w^2-8\right)+2 v^2 \left(57 w^4+8 w^2-8\right)+15 w^6+100 w^4+134 w^2  \right.\right. \\
		&\left.\left.+32\right)+4 \tilde{k}^4 \left(v^4+v^2 \left(8-25 w^2\right)-9 w^4-32 w^2-10\right)+\tilde{k}^2 \left(19 w^2-16 v^2\right)+12 \right) \ ,
	\end{aligned}
\end{equation}

\begin{equation}
	\begin{aligned}
		&\mathbb{P}^{32}_3=\mathbb{P}^{32} \left( \textbf{k}, \textbf{p}, \textbf{p}-\textbf{q} -	\textbf{k}, \textbf{q}, -\textbf{q} \right) +\mathbb{P}^{32} \left( \textbf{k}, \textbf{p}, \textbf{p}-\textbf{q} - \textbf{k},	\textbf{q}, \textbf{p}- \textbf{k}   \right) \\
		&=-\frac{k^4}{256 \tilde{k}^8 \left(\tilde{k}^2 \left(v^2+w^2-1\right)-3\right)}\left( \tilde{k}^{10} \left(v^{10}-3 v^8 \left(w^2+3\right)-15 v^6 \left(w^4-2 w^2-1\right)+v^4 \left(-9 w^6+73 w^4 \right.\right.\right. \\
		&\left.\left.\left.-77 w^2+1\right)+v^2 \left(2 w^8+28 w^6-95 w^4+82 w^2-16\right)+2 w^8-19 w^6+41 w^4-32 w^2+8\right)+\tilde{k}^8 \left(-7 v^8 \right.\right. \\
		&\left.\left.+10 v^6 \left(3 w^2+5\right)+v^4 \left(63 w^4-194 w^2-63\right)+v^2 \left(-170 w^4+202 w^2+40\right)-2 w^8+10 w^6+21 w^4-16 w^2\right.\right. \\
		&\left.\left.-20\right)+\tilde{k}^6 \left(15 v^6-v^4 \left(81 w^2+53\right)+v^2 \left(-39 w^4+310 w^2-16\right)+9 w^6-53 w^4+8 w^2-40\right)+\tilde{k}^4 \left(-13 v^4 \right.\right.  \\
		&\left.\left.+v^2 \left(46 w^2-24\right)-9 w^4+32 w^2+96\right)+8 \tilde{k}^2 \left(2 v^2+w^2-4\right)-12 \right) \ ,
	\end{aligned}
\end{equation}

\begin{equation}
	\begin{aligned}
		&\mathbb{P}^{24}_1=\mathbb{P}^{24}	\left( \textbf{k}, \textbf{p}, - \textbf{p}, \textbf{q}, - \textbf{q}	\right) +\mathbb{P}^{24} \left( \textbf{k}, \textbf{p}, -	\textbf{p}, \textbf{q}, \textbf{q} - \textbf{p} \right) \\
		&=\frac{k^4}{128 \tilde{k}^{10} v^2}\left(\tilde{k}^4 \left(v^4+6 v^2+1\right)-2 \tilde{k}^2 \left(v^2+1\right)+1\right) \left(\tilde{k}^6 v^2 \left(v^4+v^2 \left(8 w^2-2\right)+8 w^4-8 w^2+1\right) \right.  \\
		&\left.+\tilde{k}^4 \left(-6 v^4+v^2 \left(6-24 w^2\right)+4\right)+\tilde{k}^2 \left(9 v^2-8\right)+4\right) \ ,
	\end{aligned}
\end{equation}

\begin{equation}
	\begin{aligned}
		&\mathbb{P}^{24}_2=\mathbb{P}^{24} \left( \textbf{k}, \textbf{p}, \textbf{q} -	\textbf{k}, \textbf{q}, \textbf{q} - \textbf{p} \right) +\mathbb{P}^{24} \left( \textbf{k}, \textbf{p}, \textbf{q} - \textbf{k},	\textbf{q}, - \textbf{k} + \textbf{p} \right) \\
		&=\frac{k^4}{128 \tilde{k}^{10} w^2}\left( \tilde{k}^{10} w^2 \left(8 v^8+8 v^6 \left(w^2-3\right)+v^4 \left(w^4-2 w^2+25\right)+2 v^2 \left(3 w^4-6 w^2-5\right)+w^4+6 w^2+1\right)\right. \\
		&\left.-2 \tilde{k}^8 \left(20 v^6 w^2+v^4 \left(11 w^4-27 w^2-2\right)+v^2 \left(w^6+4 w^4-5 w^2+4\right)+w^6+9 w^4+10 w^2-2\right)+\tilde{k}^6 \left(v^4 \left(65 w^2 \right.\right.\right. \\
		&\left.\left.\left.-8\right)+v^2 \left(20 w^4-22 w^2+8\right)+w^6+18 w^4+46 w^2+16\right)+\tilde{k}^4 \left(4 v^4+v^2 \left(8-42 w^2\right)-2 \left(3 w^4+18 w^2+20\right)\right)\right. \\
		&\left.+\tilde{k}^2 \left(-8 v^2+9 w^2+16\right)+4 \right) \ ,
	\end{aligned}
\end{equation}

\begin{equation}
	\begin{aligned}
		&\mathbb{P}^{24}_3=\mathbb{P}^{24} \left( \textbf{k}, \textbf{p}, \textbf{p}-\textbf{q} -	\textbf{k}, \textbf{q}, -\textbf{q} \right) +\mathbb{P}^{24} \left( \textbf{k}, \textbf{p}, \textbf{p}-\textbf{q} - \textbf{k},	\textbf{q}, \textbf{p}- \textbf{k}   \right) \\
		&=\frac{k^4}{128 \tilde{k}^8 \left(\tilde{k}^2 \left(v^2+w^2-1\right)-3\right)}\left( \tilde{k}^{10} \left(v^{10}-v^8 \left(5 w^2+11\right)+v^6 \left(-5 w^4+8 w^2+35\right)+v^4 \left(w^6+17 w^4\right.\right.\right. \\
		&\left.\left.\left.+15 w^2-49\right)+v^2 \left(6 w^6-3 w^4-34 w^2+32\right)+w^6-9 w^4+16 w^2-8\right)-\tilde{k}^8 \left(5 v^8-2 v^6 \left(14 w^2+19\right) \right.\right.  \\
		&\left.\left.+v^4 \left(-7 w^4+70 w^2+69\right)+2 v^2 \left(w^6+23 w^4-27 w^2-24\right)+2 w^6-15 w^4+8 w^2+12\right)+\tilde{k}^6 \left(7 v^6 \right.\right.  \\
		&\left.\left.-v^4 \left(41 w^2+3\right)+v^2 \left(w^4+90 w^2-56\right)+w^6-3 w^4-32 w^2+24\right)+\tilde{k}^4 \left(-7 v^4+2 v^2 \left(9 w^2-16\right)\right.\right. \\
		&\left.\left.-3 w^4+24 w^2+16\right)+8 \tilde{k}^2 \left(v^2-2\right)-4 \right) \ ,
	\end{aligned}
\end{equation}

\begin{equation}
	\begin{aligned}
		&\mathbb{P}^{42}_1=\mathbb{P}^{42}	\left( \textbf{k}, \textbf{p}, - \textbf{p}, \textbf{q}, - \textbf{q}	\right) +\mathbb{P}^{42} \left( \textbf{k}, \textbf{p}, -	\textbf{p}, \textbf{q}, \textbf{q} - \textbf{p} \right) \\
		&=\frac{k^4}{128 \tilde{k}^{10} v^2}\left(\tilde{k}^4 \left(v^4+6 v^2+1\right)-2 \tilde{k}^2 \left(v^2+1\right)+1\right) \left(\tilde{k}^6 v^2 \left(v^4+v^2 \left(8 w^2-2\right)+8 w^4-8 w^2+1\right) \right.  \\
		&\left.+\tilde{k}^4 \left(-6 v^4+v^2 \left(6-24 w^2\right)+4\right)+\tilde{k}^2 \left(9 v^2-8\right)+4\right) \ ,
	\end{aligned}
\end{equation}

\begin{equation}
	\begin{aligned}
		&\mathbb{P}^{42}_2=\mathbb{P}^{42} \left( \textbf{k}, \textbf{p}, \textbf{q} -	\textbf{k}, \textbf{q}, \textbf{q} - \textbf{p} \right) +\mathbb{P}^{42} \left( \textbf{k}, \textbf{p}, \textbf{q} - \textbf{k},	\textbf{q}, - \textbf{k} + \textbf{p} \right) \\
		&=\frac{k^4}{128 \tilde{k}^{10} v^2}\left( \tilde{k}^{10} \left(v^6 \left(w^4+6 w^2+1\right)+2 v^4 \left(4 w^6-w^4-6 w^2+3\right)+v^2 \left(w^2-1\right)^2 \left(8 w^4-8 w^2+1\right)\right)  \right. \\
		&\left.-2 \tilde{k}^8 \left(v^6 \left(w^2+1\right)+v^4 \left(11 w^4+4 w^2+9\right)+v^2 \left(20 w^6-27 w^4-5 w^2+10\right)-2 \left(w^2-1\right)^2\right)+\tilde{k}^6 \left(v^6  \right.\right. \\
		&\left.\left.+2 v^4 \left(10 w^2+9\right)+v^2 \left(65 w^4-22 w^2+46\right)+8 \left(-w^4+w^2+2\right)\right)-2 \tilde{k}^4 \left(3 v^4+3 v^2 \left(7 w^2+6\right) \right.\right. \\
		&\left.\left.-2 \left(w^4+2 w^2-10\right)\right)+\tilde{k}^2 \left(9 v^2-8 w^2+16\right)+4  \right) \ ,
	\end{aligned}
\end{equation}

\begin{equation}
	\begin{aligned}
		&\mathbb{P}^{42}_3=\mathbb{P}^{42} \left( \textbf{k}, \textbf{p}, \textbf{p}-\textbf{q} -	\textbf{k}, \textbf{q}, -\textbf{q} \right) +\mathbb{P}^{42} \left( \textbf{k}, \textbf{p}, \textbf{p}-\textbf{q} - \textbf{k},	\textbf{q}, \textbf{p}- \textbf{k}   \right) \\
		&=\frac{k^4}{128 \tilde{k}^{10} v^2}  \left( \tilde{k}^{10} v^2 \left(v^8+2 v^6 \left(5 w^2+3\right)+v^4 \left(25 w^4+12 w^2+1\right)+v^2 \left(24 w^6-2 w^4-6 w^2\right)+8 w^8-8 w^6\right.\right.  \\
		&\left.\left.+w^4\right)-2 \tilde{k}^8 \left(5 v^8+9 v^6 \left(4 w^2+3\right)+v^4 \left(59 w^4+34 w^2+12\right)+v^2 \left(28 w^6-3 w^4-2 w^2-2\right)-2 w^4\right)\right. \\
		&\left.+\tilde{k}^6 \left(37 v^6+2 v^4 \left(85 w^2+72\right)+v^2 \left(137 w^4+56 w^2+76\right)-8 \left(w^4+2 w^2-2\right)\right)-4 \tilde{k}^4 \left(14 v^4+v^2 \left(31 w^2+25\right)\right.\right. \\
		&\left.\left.-w^4-8 w^2+4\right)+4 \tilde{k}^2 \left(5 v^2-4 \left(w^2+1\right)\right)+16 \right) \ ,
	\end{aligned}
\end{equation}

\begin{equation}
	\begin{aligned}
		&\mathbb{P}^{34}_1=\mathbb{P}^{34}	\left( \textbf{k}, \textbf{p}, - \textbf{p}, \textbf{q}, - \textbf{q}	\right) +\mathbb{P}^{34} \left( \textbf{k}, \textbf{p}, -	\textbf{p}, \textbf{q}, \textbf{q} - \textbf{p} \right) \\
		&=-\frac{k^4}{32 \tilde{k}^8}\left(\tilde{k}^4 \left(v^2-1\right)^2-2 \tilde{k}^2 \left(v^2+1\right)+1\right) \left(\tilde{k}^2 \left(v^2+1\right)-1\right) \left(\tilde{k}^2 \left(v^2+2 w^2-1\right)-3\right) \ ,
	\end{aligned}
\end{equation}

\begin{equation}
	\begin{aligned}
		&\mathbb{P}^{34}_2=\mathbb{P}^{34} \left( \textbf{k}, \textbf{p}, \textbf{q} -	\textbf{k}, \textbf{q}, \textbf{q} - \textbf{p} \right) +\mathbb{P}^{34} \left( \textbf{k}, \textbf{p}, \textbf{q} - \textbf{k},	\textbf{q}, - \textbf{k} + \textbf{p} \right) \\
		&=-\frac{k^4}{32 \tilde{k}^8}\left( \tilde{k}^8 \left(v^4 \left(w^4+6 w^2+1\right)+2 v^2 w^2 \left(w^4+2 w^2-3\right)+\left(w^2-1\right)^2 \left(2 w^2-1\right)\right)-2 \tilde{k}^6 \left(v^4 \left(w^2+1\right) \right.\right.  \\
		&\left.\left.+2 v^2 \left(2 w^4+5 w^2+1\right)+w^6+w^4-4\right)+\tilde{k}^4 \left(v^4+2 v^2 \left(5 w^2+4\right)+7 w^4+4 w^2-10\right)\right. \\
		&\left.-4 \tilde{k}^2 \left(v^2+2 w^2\right)+3 \right) \ ,
	\end{aligned}
\end{equation}

\begin{equation}
	\begin{aligned}
		&\mathbb{P}^{34}_3=\mathbb{P}^{34} \left( \textbf{k}, \textbf{p}, \textbf{p}-\textbf{q} -	\textbf{k}, \textbf{q}, -\textbf{q} \right) +\mathbb{P}^{34} \left( \textbf{k}, \textbf{p}, \textbf{p}-\textbf{q} - \textbf{k},	\textbf{q}, \textbf{p}- \textbf{k}   \right) \\
		&=-\frac{k^4}{32 \tilde{k}^8}\left( \tilde{k}^8 \left(v^8+4 v^6 w^2+v^4 \left(5 w^4-6 w^2-1\right)+2 v^2 w^2 \left(w^4-2 w^2+3\right)+w^4 \left(2 w^2-1\right)\right)-2 \tilde{k}^6 \left(4 v^6\right.\right. \\
		&\left.\left.+v^4 \left(11 w^2+3\right)+v^2 \left(8 w^4-2 w^2-2\right)+w^6+5 w^4+2 w^2+2\right)+\tilde{k}^4 \left(23 v^4+v^2 \left(38 w^2+24\right)+11 w^4  \right.\right. \\
		&\left.\left.+24 w^2+20\right)-4 \tilde{k}^2 \left(7 v^2+5 w^2+7\right)+12 \right) \ ,
	\end{aligned}
\end{equation}

\begin{equation}
	\begin{aligned}
		&\mathbb{P}^{43}_1=\mathbb{P}^{43}	\left( \textbf{k}, \textbf{p}, - \textbf{p}, \textbf{q}, - \textbf{q}	\right) +\mathbb{P}^{43} \left( \textbf{k}, \textbf{p}, -	\textbf{p}, \textbf{q}, \textbf{q} - \textbf{p} \right) =0 \ ,
	\end{aligned}
\end{equation}

\begin{equation}
	\begin{aligned}
		&\mathbb{P}^{43}_2=\mathbb{P}^{43} \left( \textbf{k}, \textbf{p}, \textbf{q} -	\textbf{k}, \textbf{q}, \textbf{q} - \textbf{p} \right) +\mathbb{P}^{43} \left( \textbf{k}, \textbf{p}, \textbf{q} - \textbf{k},	\textbf{q}, - \textbf{k} + \textbf{p} \right) = 0\ ,
	\end{aligned}
\end{equation}

\begin{equation}
	\begin{aligned}
		&\mathbb{P}^{43}_3=\mathbb{P}^{43} \left( \textbf{k}, \textbf{p}, \textbf{p}-\textbf{q} -	\textbf{k}, \textbf{q}, -\textbf{q} \right) +\mathbb{P}^{43} \left( \textbf{k}, \textbf{p}, \textbf{p}-\textbf{q} - \textbf{k},	\textbf{q}, \textbf{p}- \textbf{k}   \right) =0 \ ,
	\end{aligned}
\end{equation}

\acknowledgments
We thank Prof.~Sai Wang for the useful discussions. 
This work has been funded by the National Nature Science Foundation of China under grant No. 12075249 and 11690022, and the Key Research Program of the Chinese Academy of Sciences under Grant No. XDPB15.

\bibliography{biblio}

\end{document}